\documentclass[aps,pre,onecolumn,superscriptaddress,showpacs]{revtex4}
\usepackage{bm}
\usepackage{graphicx}
\usepackage{amssymb}

\def\dd{\mathrm{d}}
\def\pintun#1{{\int \dd {#1} \:}}
\def\simgr{\mathrel{\hbox{\rlap{\hbox{\lower4pt\hbox{$\sim$}}}\hbox{$>$}}}}
\def\simlw{\mathrel{\hbox{\rlap{\hbox{\lower4pt\hbox{$\sim$}}}\hbox{$<$}}}}

\begin{document}
\title{Free-energy model for fluid helium at high density}

\author{Christophe Winisdoerffer}\email{cwinisdo@ens-lyon.fr}
\affiliation{Theoretical Astrophysics Group,
     University of Leicester, Leicester,
     LE1 7RH, United Kingdom}
\affiliation{\'Ecole normale sup\'erieure de Lyon,
     CRAL (UMR CNRS No.\ 5574),
     69364 Lyon Cedex 07, France}
\author{Gilles Chabrier}\email{chabrier@ens-lyon.fr}
\affiliation{\'Ecole normale sup\'erieure de Lyon,
     CRAL (UMR CNRS No.\ 5574),
     69364 Lyon Cedex 07, France}

\date{\today}

\begin{abstract}
We present a semi-analytical free-energy model aimed at characterizing the
thermodynamic properties of dense fluid
helium, from the low-density atomic phase to the high-density fully ionized
regime. The
model is based
on a free-energy minimization method and includes various different
contributions
representative of
the correlations between atomic and ionic species and electrons. This model
allows the
computation of the thermodynamic properties of dense helium over an extended
range of
density and
temperature and leads to the computation of the phase diagram of dense fluid
helium,
with its various
temperature and pressure ionization contours. One of the predictions of the
model is
that pressure ionization occurs abruptly at $\rho \simgr 10$ g cm$^{-3}$,
{\it i.e.} $P\simgr 20$ Mbar,
from atomic helium He to fully ionized helium He$^{2+}$, or at least to a
strongly ionized
state,
without He$^{+}$ stage, except at high enough temperature for
temperature ionization to
become dominant. These predictions and this phase diagram provide a guide for
future
dynamical experiments or numerical first-principle calculations aimed at
studying the properties of helium at very high density, in particular its
metallization.
Indeed, the characterization of the helium phase diagram bears important
consequences
for the thermodynamic, magnetic and transport properties of cool and dense
astrophysical objects, among which the solar and the 
numerous recently discovered
extrasolar giant planets.
\end{abstract}


\maketitle

\section{\label{sec_intro} Introduction}

Within the past decade, over a hundred brown dwarfs, astrophysical bodies
not dense enough to sustain hydrogen fusion in their core, and extrasolar
giant planets, {\it i.e.} jovian planets orbiting stars outside the solar system,
have been discovered. These objects are composed essentially of hydrogen and
helium. Given their large gravity and relatively low temperature, within astrophysical
standards, the hydrogen and heluim fluid is under an atomic
or molecular form in the outermost part of the body and under the form of
a fully ionized electron-ion plasma in the innermost regions. Such an internal structure is common
to many so-called compact objects, from our own jovian planets to the external
layers of white dwarfs or neutron stars. The characterization of the structure
and cooling properties of these compact objects thus requires the knowledge of the
thermodynamic properties of dense hydrogen and helium fluids, and more importantly
a realistic description of the partial, pressure ionization regime. Given the
large variations of thermodynamic conditions characteristic of the structure and evolution
of such astrophysical bodies, these thermodynamic properties, characterized
by the equation of state (EOS), must be calculated over several orders of magnitudes in density
and temperature. As discussed below, the
necessity to calculate the thermodynamic properties
 over such a large range of
conditions precludes the use of heavy computer simulations and thus necessitates
the derivation of EOS models which allow extensive calculations within a
reasonable amount of computer time, unfortunately at the price of a more approximate,
or say phenomenological description of the properties of matter at high-density.

Interestingly enough, these EOS of dense matter under astrophysical conditions
can now be probed on Earth by shock wave experiments. Future large laser experiments,
like e.g. the NIF project at Livermore or the LMJ project in France, will reach
conditions characteristic of the deep interior of the aforementioned astrophysical
bodies. So not only the calculation
of dense matter EOS is of interest for astrophysical applications, but it is necessary
for the confrontation of theory with existing and future high-pressure experiments, yielding eventually
a correct knowledge of the properties of matter under extreme conditions.
Hydrogen, the most common element in the universe, has been studied extensively, both on the experimental 
and theoretical fronts, and the EOS of dense hydrogen becomes more and more constrained, although the 
very regime of pressure ionization still remains ill-determined. The same
cannot be said for helium. Although some experiments exist in the regime of
neutral helium at high-density, as detailed below, the regime of helium pressure
ionization, from He to He$^{+}$ and He$^{2+}$ remains for now unexplored, and no attempt
has been made to give a detailed theoretical description of these domains. It is the
very purpose of the present paper to derive an EOS for dense, partially ionized helium,
covering the gap between the
previous study of dense neutral helium \cite{aparicio94} and the fully ionized 
regime \cite{chabrier98} \cite{potekhin00}. As mentioned above, not only the calculation of such a dense helium EOS is
necessary for a description of the thermodynamic properties of astrophysical
compact objects, in particular the recently discovered gaseous exoplanets, but it provides a
useful guide
for future high-pressure shock-wave or laser experiments.

The paper is organised as follows. In Sec. \ref{sec_gal_cons}, we briefly comment on
the general formalism underlying the present calculations. The various contributions
entering our general model free-energy are presented in detail in Sec. \ref{sec_Fmodel}.
The results, and the limitations of the model are presented in Sec. \ref{sec_results}. Special
attention is devoted to the impact of various approximations in the free-energy calculation
on the final results. Section \ref{sec_concl} is devoted to the conclusion.

\section{\label{sec_gal_cons} General considerations}
\subsection{Chemical picture of a dense plasma}
Equation-of-state calculations can be divided into two generic categories.
The ``physical approach'' is formally exact as
it involves only fundamental particles, electrons and nuclei,
interacting through the Coulomb potential. The
partition function is calculated using the eigenvalues corresponding to
this $N$-body system. In practice, however, the exact solution cannot be calculated,
in particular when bound-states form, and either perturbative expansions
or approximate numerical schemes must be used. The validity of the expansions is limited to high temperatures
and/or low-densities, {\it i.e.} apply to weakly or moderately coupled plasmas. The regime
of pressure ionization thus cannot be described by such expansion schemes. Numerical
technics, such as density functional theory, molecular dynamics or path-integral Monte Carlo simulations,
do extend to the strongly correlated regime but the description of the pressure
ionization regime then becomes a formidable 
task, and involves also physical approximations in the calculations of
either the electron functional or the nodal functions, not mentioning 
the finite size effects due to the
limited number of particles in the simulation. 
In practice, these simulations do not allow
the calculation of thermodynamic quantities over a large
range of temperatures and densities, as needed for practical applications, as mentioned earlier. 
For this reason, a more phenomenological approach has been developped which combines
a simplified description of the properties of dense matter and a semi-analytical derivation,
allowing the calculations of extended thermodynamic tables with moderate computer time investment.
This is the so-called ``chemical picture''. In this approach, the
basic particles are no longer only electrons and nuclei but also bound species (atoms, molecules, ions),
which are characterized by their interparticle potentials. That
means that the particles remain distinguishable (in a
classical sense) in the
plasma, with their own identities and interaction properties. The problem thus
reduces to the free-energy minimization
of a multi-component system, taking into account chemical and ionization equilibrium
between the various species.
Although certainly of doubtful validity in the regime of pressure ionization, where the concepts
of pair potential and bound-states become meaningless, this approach has been shown to
yield reasonably accurate descriptions of hydrogen at high density 
\cite{saumon92} \cite{saumon00}.
Moreover, as mentioned above, this approach presents the advantage of being
semi-analytical and thus has a precious practical interest for EOS calculations.
Last but not least, the chemical approach offers the noticeable advantage of clearly identifying the terms
and the approximations aimed at describing various physical effects. Such terms can be added or
removed with limited effort, allowing a rapid identification of the dominant contributions responsible
for the thermodynamic properties of matter under complex conditions.
Therefore, despite its shortcomings, the chemical 
approach should be seen as a useful alternative to the ``exact" physical approach.

\subsection{General free-energy model}
The chemical approach is based on the minimization of the free-energy
$F(\{N_i\},T,V)$ corresponding to a system containing $\{N_i\}$ 
different species
inside a volume $V$ at temperature $T$.
This minimization $\delta F=\sum_i\frac{\partial F}{\partial N_i}\delta N_i=0$
must satisfy the electroneutrality condition and the st\oe chiometric
conditions corresponding in our case to the following set of chemical equations:
\begin{eqnarray}
\left\{
\begin{array}{lcl}
{\mathrm {He}} & \leftrightharpoons & {\mathrm {He}}^+ + {\mathrm e}^- ,\\
{\mathrm {He}}^+ & \leftrightharpoons & {\mathrm {He}}^{2+} + {\mathrm e}^-
.
\end{array}
\right.
\end{eqnarray}
The canonical partition function of the system
$\mathcal{Z}$ is assumed
to be factorizable into different contributions, so that
 the free-energy $F=-kT\ln \mathcal{Z}$ can be split into the sum of translational,
configurational and internal contributions \cite{graboske69} \cite{fontaine77}.
Adding up the correction arising from the quantum behaviour
of the heavy particles, one gets:
\begin{eqnarray}
F(\{N_i\},T,V)&=&F_{\mathrm{id}}(\{N_i\},T,V)+
F_{\mathrm{conf}}(\{N_i\},T,V) \nonumber \\
&+&F_{\mathrm{int}}(\{N_i\},T,V)+
F_{\mathrm{qm}}(\{N_i\},T,V)
.\nonumber \\
& &
\label{eq_sepF}
\end{eqnarray}
The conditions of validity of such a separability are:
\begin{itemize}
\item the discretization of the eigenvalues corresponding to the 
translation degrees of freedom and to the center-of-mass positions are
negligible. This is the quasi-classic approximation;
\item there is no coupling between the translation degrees of freedom and
the center-of-mass positions;
\item the internal energy levels remain essentially unperturbed by
the interactions with surrounding particles.
\end{itemize}
If the two first conditions are satisfied in the present context, the last one certainly becomes invalid in the pressure ionization regime.
We expect this regime, however, to cover a limited range of density, as
pressure ionization generally occurs rather abruptly.
Eventually, only comparison with experimental data can give a quantitative estimate of the discrepancy
due to this underlying factorization condition.
The various contributions to $F$ are described in the next section.

\section{\label{sec_Fmodel} Free-energy model}
We first present the models used to calculate the contributions 
to the total free-energy
arising from each different species, He, He$^{+}$, He$^{2+}$. Then, we describe the modelization
of the coupling between these various species.
\subsection{\label{sec_Fmodel_He0} Model for atomic helium He}

\subsubsection{The kinetic free-energy $F_{\mathrm{id}}$}
The ideal part of the free-energy, corresponding to the kinetic part of the 
Hamiltonian, is given by \cite{landau9}:
\begin{equation}
F_{\mathrm{id}}(N,T,V)=-Nk_BT\left[1+\ln \left(\frac{V}{N}
\left(\frac{2\pi Mk_BT}{h^2}\right)^{3/2}\right)\right]
,
\end{equation}
where $N$ is the number of helium atoms of mass $M$ inside the volume $V$
at temperature $T$.
 
\subsubsection{\label{sec_Fconf} The configurational free-energy 
$F_{\mathrm{conf}}$}
\begin{figure}
\includegraphics[width=\columnwidth]{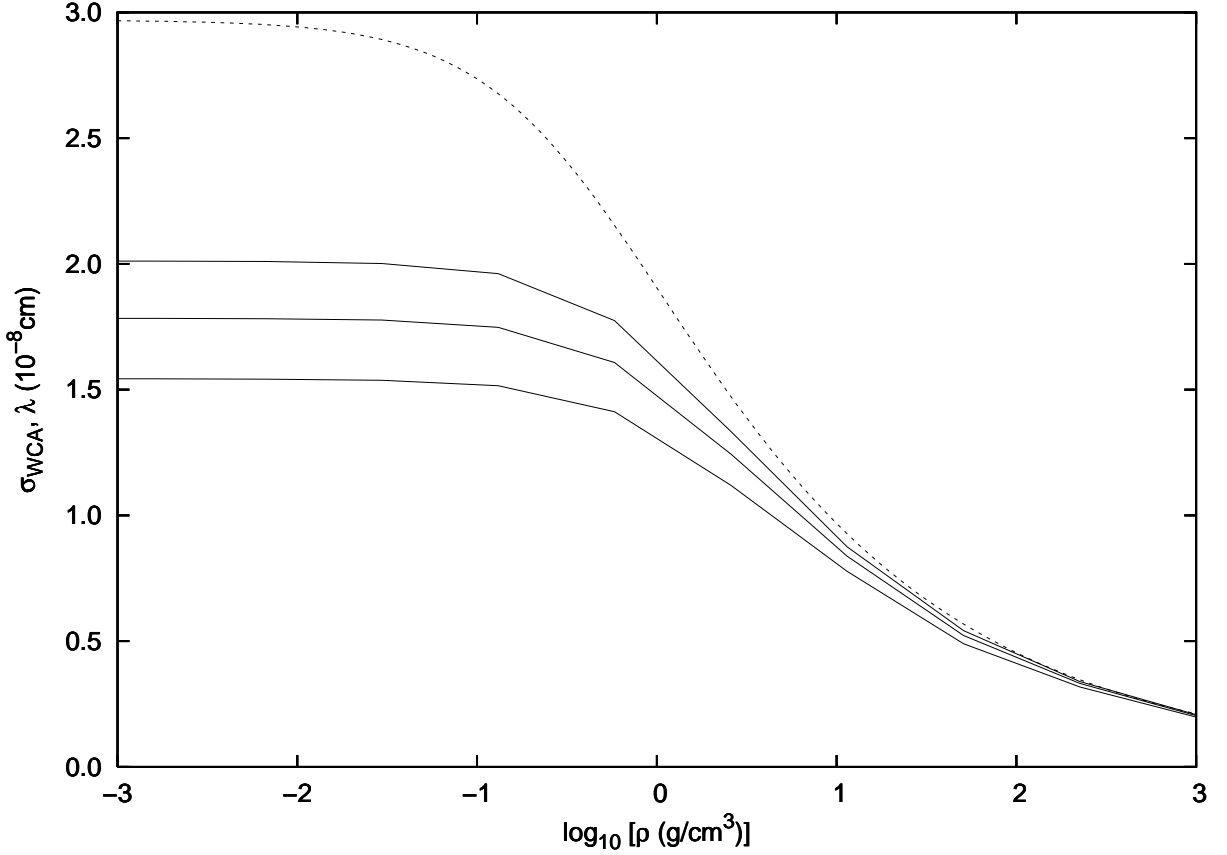}
\caption{\label{fig_sigma} Hard-sphere radii
$\sigma_{\mathrm{WCA}}$ for atomic helium
(solid line), for $T=10^{3.0}$ K, $T=10^{3.5}$ K and $T=10^{4.0}$ K
from top to bottom, 
and break-point $\lambda$ (dotted line) as a function of the density.}
\end{figure}

\begin{figure}
\includegraphics[width=\columnwidth]{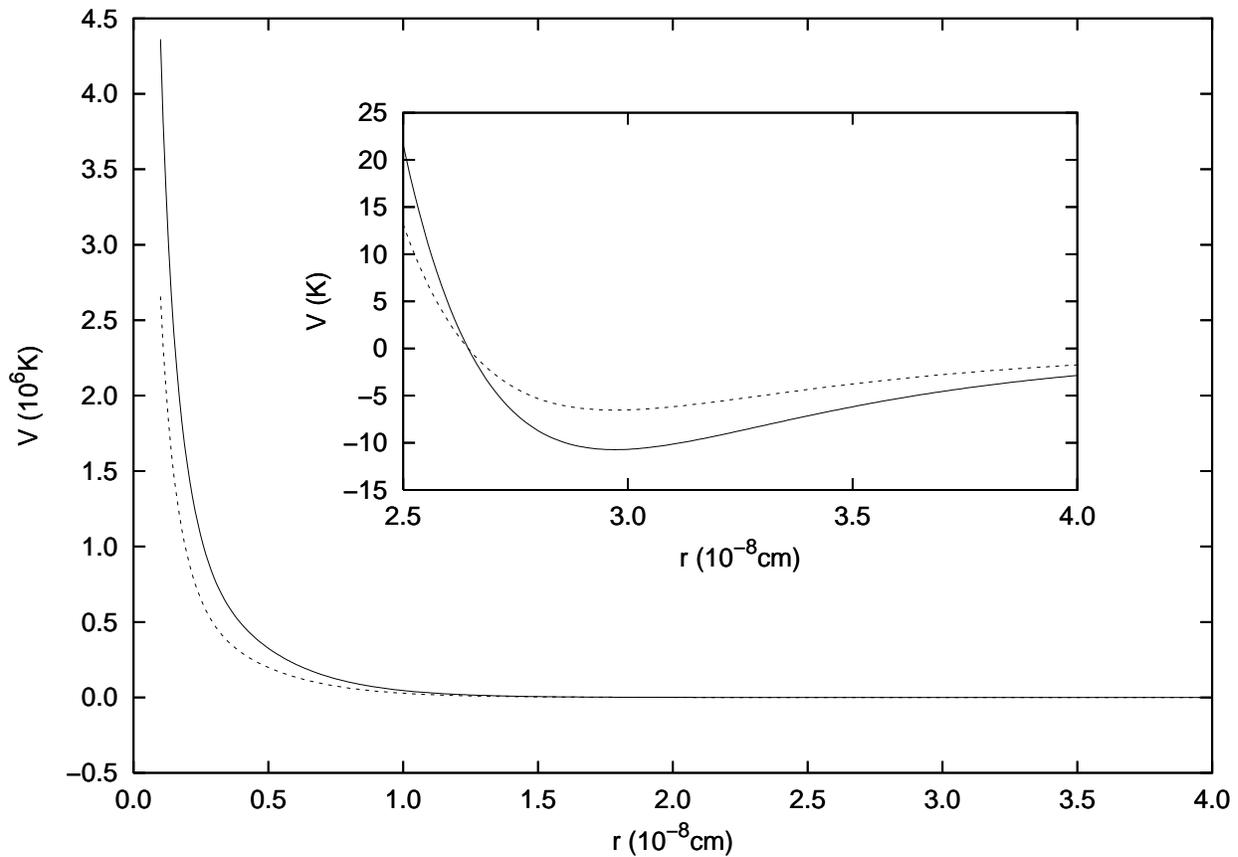}
\caption{\label{fig_aziz} Interaction potential 
between two helium atoms, without $N$-body correction
(solid line) and with the $N$-body softening correction at $\rho=10$ g/cm$^3$ (dotted
line).}
\end{figure}

\begin{figure}
\includegraphics[width=\columnwidth]{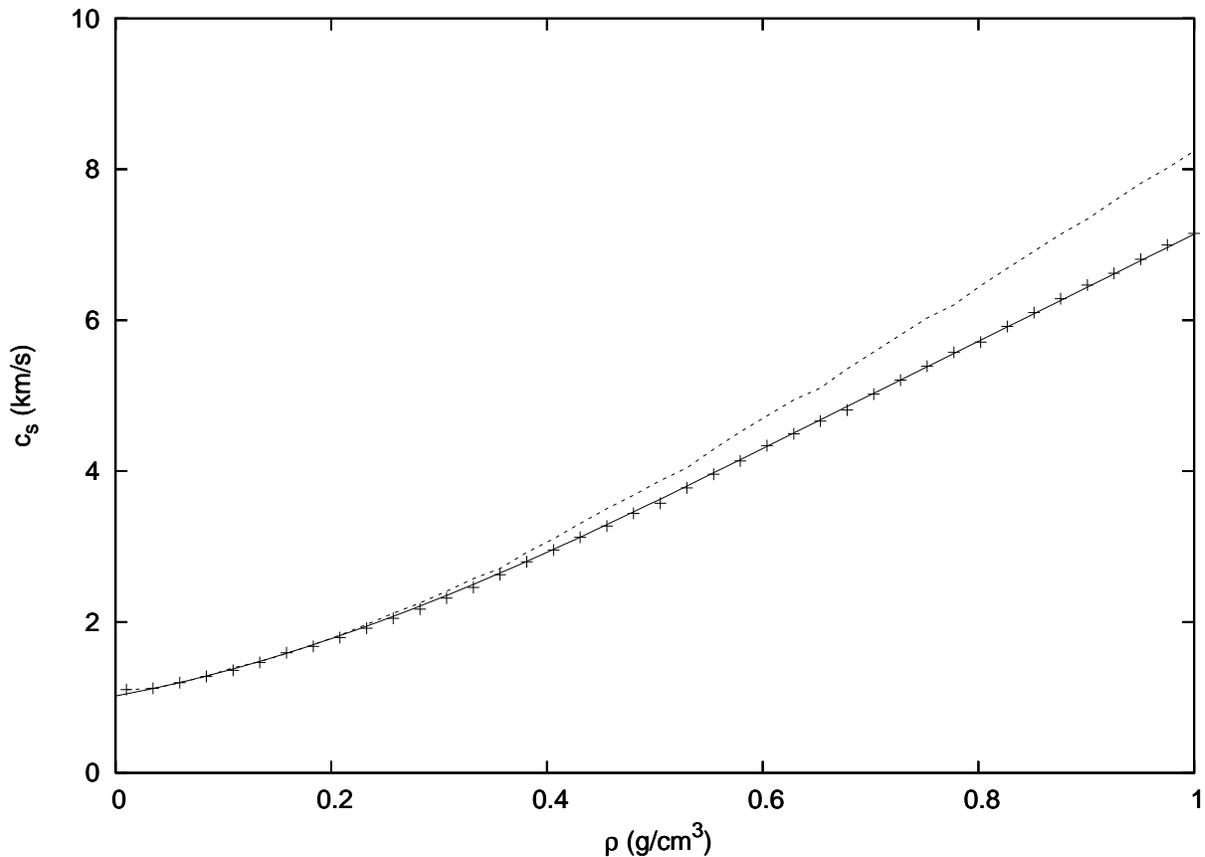}
\caption{\label{fig_sound} Comparison between the experimental measures of 
the adiabatic sound velocity  \cite{letoullec89} (solid line) 
and the present calculations with $(C,D)=(0.44,0.8$ cm$^3$/g$)$ (crosses) and
$(C,D)=(0,0$ cm$^3$/g$)$ (dotted line).}
\end{figure}

\begin{figure}
\includegraphics[width=\columnwidth]{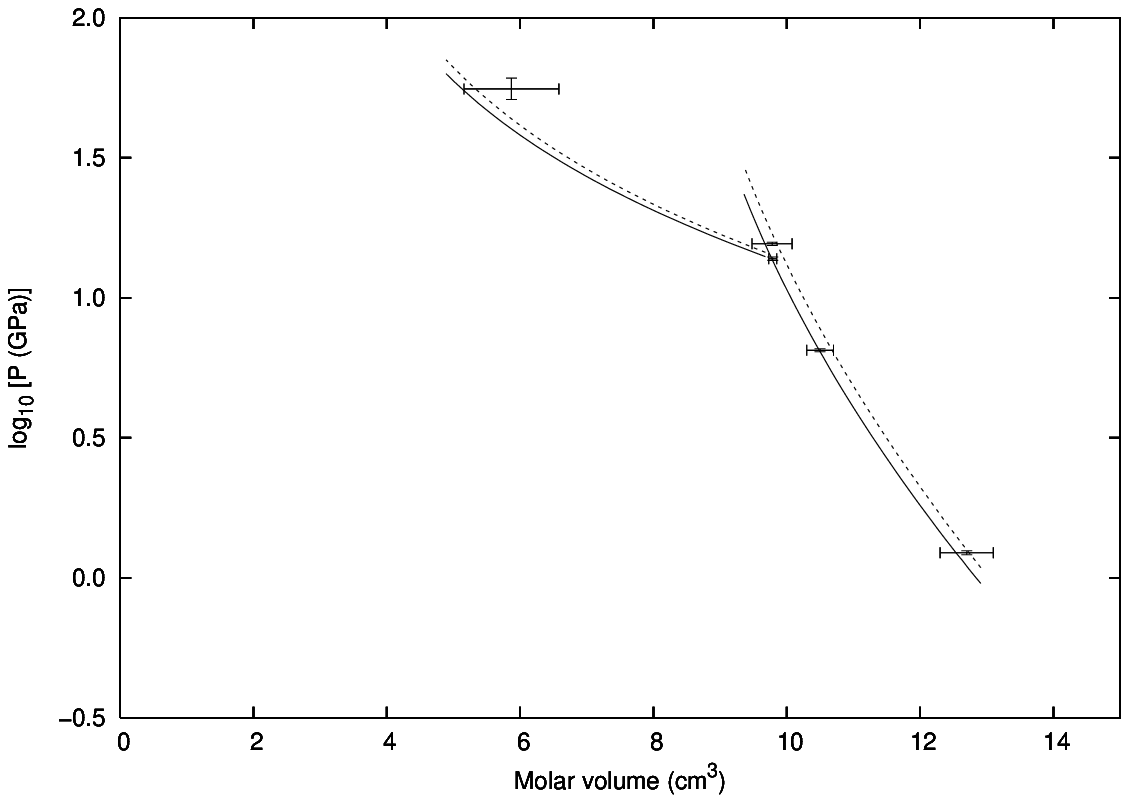}
\caption{\label{fig_shock} Comparison between the experimental
single and double-shock Hugoniot curves \cite{nellis84}
 and the present calculations with $(C,D)=(0.44,0.8$ cm$^3$/g$)$ (solid line)
and $(C,D)=(0,0$ cm$^3$/g$)$ (dotted line). }
\end{figure}

The configurational free-energy $F_{\mathrm{conf}}$, arising from the interactions between helium atoms,
is calculated within the Weeks-Chandler-Andersen \cite{weeks71a}
\cite{weeks71b}
(WCA) perturbation theory. The interaction potential
$\Phi(r)$ is split into a reference potential 
$\Phi_{\mathrm{ref}}(r)$ and a perturbative part $\Phi_{\mathrm{pert}}(r)$.
Truncating the perturbative expansion of the free-energy
after the first order, the so-called
high-temperature approximation (HTA), yields:
\begin{equation}
F_{\mathrm{conf}}=F_{\mathrm{ref}}(T,V,N)
+\frac{N^2}{2V}\pintun{\bm r} \Phi_{\mathrm{pert}}(r)
g_{\mathrm{ref}}(r,V,N)
.
\label{eq_HTA}
\end{equation}
The problem is thus reduced to the potential separation and to 
the calculation of $F_{\mathrm{ref}}(T,V,N)$ and $g_{\mathrm{ref}}(r,V,N)$.
Concerning the first point, we use a modification of the procedure of
Kang {\it{et al.}} \cite{kang85}, namely:
\begin{equation}
\begin{array}{lclr}
\Phi_{\mathrm{ref}}(r) & = &
\left\{
\begin{array}{ll}
\Phi(r)-\left(\Phi(\lambda)+\frac{\dd \Phi}{\dd r}|_{r=\lambda}(r-\lambda)
\right) &
{\mathrm {if}} \;\;\; r< \lambda ,\\
0       & {\mathrm {if}} \;\;\; r\geq \lambda ,\\
\end{array}
\right. \\
\Phi_{\mathrm{pert}}(r) & = &
\left\{
\begin{array}{ll}
\Phi(\lambda)+\frac{\dd \Phi}{\dd r}|_{r=\lambda}(r-\lambda) &
{\mathrm {if}} \;\;\; r< \lambda ,\\
\Phi(r)       & {\mathrm {if}} \;\;\; r\geq \lambda ,\\
\end{array}
\right.
\end{array}
\end{equation}
where $\lambda=(a_{\mathrm{fcc}}^{-3}+r^{\ast-3})^{-1/3}$;
$a_{\mathrm{fcc}}=(\sqrt{2}/(N/V))^{1/3}$ and $r^{\ast}$ 
corresponds to the minimum of the potential $\Phi(r)$.
This choice for
the density-dependent break-point $\lambda$ has the advantage to give a continuously
differentiable $\lambda$.
Concerning the second point,
we approximate the repulsive reference potential by a hard-sphere potential. The
 hard-sphere radius $\sigma$ is calculated from the Barker-Henderson criterion:
\begin{equation}
\sigma_{\mathrm{BH}}=\int_0^{\infty} \dd r \: (1-e^{-\beta 
\Phi_{\mathrm{ref}}})=
\int_0^{\sigma_{\mathrm{BH}}} \dd r \: (1-e^{-\beta \Phi_{\mathrm{ref}}})
,
\end{equation}
with the Verlet \& Weiss correction \cite{verlet72}
to include a density-dependence:
\begin{equation}
\sigma=\sigma_{\mathrm{BH}}\left(1+\frac{\sigma_1}{2\sigma_0}\delta\right),
\end{equation}
where $\delta$ is a function of the temperature and $\sigma_1/2\sigma_0$
is a function of T and $\sigma$. This non-linear equation is solved by direct iteration, 
using $\sigma_{\mathrm{BH}}$ as an initial guess for $\sigma\equiv\sigma(T,n)$.
An example of the evolution of $\sigma$ with density and
temperature is presented in Fig. \ref{fig_sigma}.
The free-energy and the radial distribution functions for the hard-sphere reference system 
$F_{\mathrm{ref}} \equiv F_{\mathrm{HS}}$, $g_{\mathrm{ref}} \equiv g_{\mathrm{HS}}$, are obtained
analytically \cite{mansoori71} \cite{grundke72}.\\
To describe the 
interaction between two helium atoms, we choose the
Aziz \& Slaman \cite{aziz91} potential for $r\geq 1.8$ \AA, and the
Ceperley \& Partridge \cite{ceperley86} one for $r< 1.8$ \AA. 
Following Aparicio \& Chabrier \cite{aparicio94}, this two-body potential is modified by a density-dependent function
to mimic the softening due to $N$-body effects at high density:
\begin{equation}
\Phi(r)=\left((1-C)+\frac{C}{1+D\rho}\right)\Phi_{\rho \rightarrow 0}(r)
.
\end{equation}
The two parameters $C$ and $D$ are optimized to reproduce
the experimental measures of adiabatic sound velocity \cite{letoullec89}. A $\chi^2$ minimization yields $(C,D)=(0.44,0.8$ cm$^3$/g$)$.
This potential is illustrated in Fig. \ref{fig_aziz} whereas 
Fig. \ref{fig_sound} compares
the measured sound velocity and the one calculated with our potential. 
Fig. \ref{fig_shock} compares the Hugoniot curves obtained with the present atomic helium
free-energy model and interatomic potential with presently available
shock-wave experiments \cite{nellis84}.
These comparisons assess the validity of the present model at least
up to the limit of the data, {\it i.e.} P$\simeq 1$ Mbar.


\subsubsection{The internal free-energy $F_{\mathrm{int}}$}
The divergence of the internal partition function, $\sum_l g_l \exp(-E_l/k_BT)$,
of an isolated atom is a well-known problem in statistical physics. It emphasises the
necessity to take into account the interactions between atoms in the calculation of 
the internal partition function, ${\cal Z}_{\mathrm{int}}=\exp(-\beta F_{\mathrm{int}})$.
For a density $n$, each atom has a typical available volume $n^{-1/3}$
so that, as density increases, the levels
associated with the highest eigenvalues will move into the continuum. When the density is high
enough to
disturb even the ground-state, the electrons can no longer
remain bound to the nuclei: this is the pressure ionization phenomenon. 
We have included
the effect of the interactions of surrounding particles on the internal partition function
of helium within the so-called occupation 
probability formalism \cite{hummer88} (OPF).
The OPF ensures the statistical-mechanical 
consistency between the configurational free-energy characterizing 
the interactions
between atoms, $F_{\mathrm{conf}}$,
and the internal free-energy contribution, $F_{\mathrm{int}}$. The OPF has been extensively
presented in various papers (see e.g.~\cite{saumon91}), 
and is only briefly outlined for completeness.
We consider a system of interacting particles, of free-energy
$F=F_{\mathrm{id}}-k_BT\ln {\cal Z}_{\mathrm{int}} +f$, where
$f$ is the non-ideal term. Within the OPF, the total free-energy can be 
rewritten under the form:
\begin{equation}
F=F_{\mathrm{id}}-k_BT\ln \tilde {\cal Z} +f-\sum_{\alpha}N_{\alpha}
\frac{\partial f}{\partial N_{\alpha}}
\label{eq_FconfHM}
,
\end{equation}
with
\begin{equation}
\tilde {\cal Z}=\sum_{\alpha}\omega_{\alpha}g_{\alpha}
e^{-\beta E_{\alpha}} \;\;\;\;\;\;\; {\mbox {and}} \;\;\;\;\;\;\;
\omega_{\alpha}=\exp \left( -\beta \frac{\partial f}
{\partial N_{\alpha}} \right)
.
\end{equation}
The term $\omega_{\alpha}$ can be seen as the probability that the eigenstate $\alpha$ of
the atom
still exists in the midst of
the surrounding particles. These factors 
$\omega_{\alpha}$ are calculated consistently from the configurational term
$f$, and the term $\sum_{\alpha}N_{\alpha} 
\frac{\partial f}{\partial N_{\alpha}}$ ensures the statistical-mechanical
consistency (see \cite{hummer88}). The OPF has several noticeable advantages, among which:
\begin{itemize}
\item $\omega_{\alpha}$ decreases monotoneously and continuously with increasing density,
 ensuring the convergence of $\tilde{\cal Z}_{\mathrm{int}}$ and 
the derivability of $F_{\mathrm{int}}$;
\item no ill-controlled energy shifts of the levels are introduced, as required
from the condition of factorizability of the partition function (Eq. \ref{eq_sepF}).
Experiments at low-density \cite{wiese72} and calculations
\cite{grabowski87} \cite{seidel95} do not show such energy shifts;
\item the probabilistic interpretation of $\omega_{\alpha}$ enables us
to combine several occupation probabilities arising from statistically
independent interactions. We will come back to this point in Sec. 
\ref{sec_coupling}.
\end{itemize}
The exact solution, in principle, requires the knowledge of all the
interaction potentials between an atom in state $\alpha$ and an other
one in state $\alpha'$. In the absence of such information,
we have adopted the simplest
approach which consists to characterize excited state interactions by hard-sphere excluded volumes in the phase space.
The hard-sphere radii are calculated with the scaling law
derived by Aparicio \& Chabrier \cite{aparicio94} (Eq. (14a) and
(14b)). 
Within the first order in the expansion of the non-ideal part $f$ of the free-energy (Eq. \ref{eq_HTA}),
the $\omega_{\alpha}^{\mathrm{HS}}$ for the excited states are thus given by:
\begin{equation}
\omega_{\alpha}^{\mathrm{HS}}=\exp \left( -\beta \frac
{\partial f_{\mathrm{HS}}(\{N_{\alpha}\},V,T)}{\partial N_{\alpha}} \right)
.
\label{eq_omegaHS}
\end{equation}
This nonlinear equation is solved iteratively by using results obtained
within the low excitation approximation (LEA) and low density approximation
(LDA), $\omega_{\alpha}^{\mathrm{HS,LEA+LDA}}=\exp(-\pi N (\sigma_{\alpha}
+\sigma_1)^3/6V)$, as initial guess.

\subsubsection{\label{sec_Fqm} The quantum correction of the free-energy 
$F_{\mathrm{qm}}$}
We have taken into account the correction to the free-energy arising from quantum effects due to the finite size
of the atoms by keeping the first order of the Wigner-Kirkwood
$\hbar^2$ expansion of ${\mathrm {Tr}}\; [e^{-\beta H}]$ \cite{wigner32}
\cite{kirkwood33}
\cite{landau9}:
\begin{eqnarray}
F_{\mathrm{qm}}=\frac{\hbar^2}{24k_BTVM_{\mathrm{He}}}
N^2\pintun{{\bm r}} {\bm \nabla}^2\Phi(r)
g(r)
.
\end{eqnarray}
$\Phi(r)$ corresponds to the potential explicited in Sec. \ref{sec_Fconf},
and $g(r)\equiv y(r)e^{-\beta \Phi(r)}$ is approximated by 
$y_{\mathrm{HS}}(r)e^{-\beta \Phi(r)}$.

\subsection{Model for the partially ionized plasma \{He$^{+}$, e$^-$\}}
Because of the presence of bound-states, the treatment of He$^{+}$ presents the
same difficulties as for He.
We adopt the same formalism, namely
the WCA perturbation expansion, to calculate the He$^{+}$ configurational free-energy
(with the hard-sphere model as the reference system) and the OPF
to treat the internal partition function.
For the long-range interaction potential between He$^{+}$ ions, we take a
Yukawa potential, $e^{-k_{\mathrm s}r}/r$, where the density- and temperature-dependent screening wave vector
is given by \cite{chabrier90}:
\begin{equation}
k_{\mathrm s}(n,T)=\frac{1}{\sqrt{2}} k_{\mathrm{TF}}
[\theta^{1/2} F_{-1/2}(\mu/k_BT)]^{1/2}
,
\end{equation}
where $k_{\mathrm{TF}}=(4m_ee^2/\pi\hbar^2)^{1/2}(3\pi^2 n_e)^{1/6}$ is the
Thomas-Fermi screening wave vector, $n_e$ is the total free electron density, $\theta=T/T_{\mathrm F}$ is the
electronic degeneracy parameter ($T_{\mathrm F}$ is the electron Fermi temperature),
$F_{n}$ is the Fermi integral of order $n$, and $\mu/k_BT$ is the electron chemical potential
defined by $F_{1/2}(\mu/k_BT)=2\theta^{-3/2}/3$. \\
For the treatment of the internal free-energy, we need a 
scaling law to associate a hard-sphere radius to
the excited states of He$^{+}$. Since He$^{+}$ is hydrogen-like, and the
energy levels are degenerate toward the orbital quantum number $l$, we write this scaling law as:
\begin{equation}
\sigma_n=n^2 \sigma_1
,
\end{equation}
where $\sigma_1$ is the WCA hard-sphere radius associated to the
ground-state, and $n$ is the main quantum number.\\
The calculations then proceed exactly as in Sec. \ref{sec_Fmodel_He0}.

\subsection{\label{sec_fully} Model for the fully ionized plasma 
\{He$^{2+}$, e$^-$\}}
The free-energy of a fully ionized electron-ion plasma (FIP)
has been calculated by Chabrier \& Pothekin \cite{chabrier98} and Potekhin
\& Chabrier \cite{potekhin00}. These authors
provide analytical
parametrizations for the various thermodynamic quantities.
We refer the reader to these papers for a description of the fully ionized plasma model.

\subsection{\label{sec_coupling} Interactions between different species}
Besides all the aforedescribed contributions to the free-energy, arising from
interactions between species of same nature, we must also 
include contributions arising from the interactions between species
of {\it different} nature.
\subsubsection{Hard-sphere interactions between atoms and ions}
The first order interaction between the atomic and ionic species He, He$^+$ and He$^{2+}$
is the hard-sphere excluded volume interaction,
$F_{\mathrm{HS}}(\{N_{\mathrm{He},\alpha},N_{\mathrm{He}^+,\alpha},
N_{\mathrm{He}^{2+}}\},
\{\sigma_{\mathrm{He},\alpha},\sigma_{\mathrm{He}^+,\alpha},
\sigma_{\mathrm{He}^{2+}}\},V,T),$
with a radius $\sigma_{\mathrm{He}^{2+}}\equiv 0$ for the He$^{2+}$ ions,
calculated consistently from the hard-sphere free-energy of a multicomponent interacting system \cite{mansoori71}.
It can be shown easily that the contribution arising from the 
$\sigma_{\mathrm{He}^{2+}}=0$ component
is equivalent to renormalizing the ideal (kinetic) term for this species with a volume
$V^\prime=(1-\eta)V$, where $\eta=\sum_{i\in\{\mathrm{HS}\}}
\pi n_i\sigma_i^3/6$ corresponds to the total packing fraction \cite{saumon92}.
This term thus takes into account the He-He, He$^+$-He$^+$, 
He-He$^+$, He-He$^{2+}$ and He$^+$-He$^{2+}$ interactions.
Note that, contrarily to previous approaches, we do not consider excluded volume interactions between bound 
species and free electrons. Indeed, such an approach does not seem to be
justified, for the quantum exclusion principle applies only to electrons in the same state. The entire
volume of the system is thus available to the majority of the free electrons, even in the presence of
bound species, as far as the free electrons are in a quantum state
different from those corresponding to the bound-states. 
In any event, we have checked
that the introduction of an excluded volume for the electrons does not modify significantly the final
results. 

\subsubsection{Induced interactions between atoms and ions}
The presence of charges in the neighbourhood of species with bound-states
has two consequences. The first one is the induced polarization
due to the electronic cloud, which translates into a related contribution to the free-energy.
The second one is the induced Stark effect on the bound-states,
due to the ambient electric field which modifies the one associated to the atom nucleus.
These two effects have been taken into account in our model as described below.\\
\paragraph{Polarization effects\\}
The polarization contribution to the free-energy arising from the 
interactions between the charges and the neutral atoms He has been
handled as in previous $N$-body approaches
\cite{ebeling88} \cite{saumon92}:
\begin{equation}
F_{\mathrm{pol}}=\frac{2k_BT}{V}N_{\mathrm{He}}
\sum_{i=\mathrm{He}^+,\mathrm{He}^{2+},e^-}
N_iB_{\mathrm{He},i}
.
\end{equation}
The second virial coefficients $B_{\mathrm{He},i}$ are given by:
\begin{equation}
B_{\mathrm{He},i}=2\pi\int_{\sigma_{\mathrm{He-i}}}^\infty \dd r\, r^2 (1-e^{-\beta \Phi^i_{\mathrm{pol}}}) 
,
\end{equation}
where
\begin{equation}
\Phi^i_{\mathrm{pol}}(r)=-\frac{Z_ie^2\alpha_i}{2}
\left[\frac{1+k_{\mathrm {s}}r}
{r^2+\sigma_{\mathrm{He-i}}^2}\right]^2 \exp\left(-2k_{\mathrm {s}}r
\right)
\end{equation}
is the polarization potential between He and the species $i$. The two 
free parameters $\sigma_{\mathrm{He-i}}$ and $\alpha_i$ are
the hard core radius and the polarizability. For the
He-He$^{2+}$ and He-e$^-$ interactions, the hard core 
radius is chosen to be the He atom ground-state radius, $\sigma_{\mathrm{He-He}}^{\mathrm{HS}}$, 
and the polarizability (which has the dimension of a volume) is equal 
to $(\sigma_{\mathrm{He-He}}^{\mathrm{HS}})^3$. For the He-He$^{+}$ interaction, the hard core
radius is $\sigma_{\mathrm{He-He}^+}^{\mathrm{HS}}
= (\sigma_{\mathrm{He-He}}^{\mathrm{HS}}+
\sigma_{\mathrm{He}^+\mathrm{-He}^+}^{\mathrm{HS}})/2$ and
the polarizability is equal to $(\sigma_{\mathrm{He-He}^+}^{\mathrm{HS}})^3$.
\medskip

\paragraph{Electric microfield effects\\}
Stark effect on the bound-states, arising from the electric microfield $\bf E$ due to the
surrounding charges, is also treated within the framework of the OPF.
The occupation probability associated with the Stark interaction on the internal states
of He and He$^{+}$ is given by \cite{hummer88}:
\begin{equation}
\omega_{\alpha}^{\mu \mathrm{E}}=\int_0^{\beta_{\alpha}^{\mathrm {crit}}}\dd \beta
{\cal P}(\beta)
,
\label{eq_omegamicroE}
\end{equation}
where $\beta=(4\pi\varepsilon_0 a^2/Ze) E$
is the dimensionless
electric field ($Ze$ is the ion charge and $a=(4\pi n/3)^{-1/3}$ is the
mean interparticle distance), ${\cal P}(\beta)$ is the probability that the central
ionic center experiences a field between $\beta$ and $\beta+\dd \beta$, 
and $\beta_{\alpha}^{\mathrm {crit}}$ is a critical field associated to each
bound-state $\alpha$. Potekhin {\it{et al.}} \cite{potekhin02} 
have calculated the microfield distribution of an atom (neutral ionic center)
or an ion (charged ionic center) immerged in a surrounding ionized plasma.
These calculations take into account the interactions in the plasma ($\Gamma=(Ze)^2/akT\ne 0$), and
recover the Holtzmark limit in the case of a non-interacting, perfect gas ($\Gamma=0$).
These authors provide analytical formulae for
$Q(\beta, \Gamma)=\int_0^{\beta} \dd t {\cal P}(t, \Gamma)$ in the
case of a neutral or a charged central ionic center.
Note that $Q(\beta,\Gamma)$ and thus the probability $\omega_{\alpha}^{\mu \mathrm{E}}$
not only depend on the temperature, as in the Holtzmark limit, but depend also on the density,
through the parameter $\Gamma$. The critical fields
are given by Hummer et Mihalas \cite{hummer88} in the case of a
hydrogen-like system. We have directly applied their prescription to
He$^+$, and used the similarity between a He atom and a hydrogen-like
system, with a central charge equal to $7/4$ for the ground-state and $1$
for the ${\it 1snl}$-type levels \cite{aparicio94}, to
calculate the critical fields corresponding to atomic helium He.

\subsubsection{\label{sec_FHe1He2} Long range interaction between 
He$^+$ and He$^{2+}$}
The remaining coupling contribution between the various species stems from the
long range Coulomb interaction between helium ions He$^+$, He$^{2+}$ and electrons.
Short distance interactions due to the internal levels of He$^+$ have been considered in
the previous sections. The treatment of the long range Coulomb interaction between the two
ionic species will certainly have some impact in the pressure 
ionization regime where He$^+$ and He$^{2+}$ coexist, but
will not modify the rest of the phase diagram.
This contribution, however, is difficult to evaluate accurately. 
Considering the He$^+$-He$^{2+}$ interaction as a pure Coulomb contribution,
thus representing the He$^+$-He$^{2+}$ fluid as an interacting two-component $Z_1=1$,
$Z_2=2$ point-charge plasma is not satisfactory, for it precludes a correct
treatment of the internal levels of He$^+$, which has been included in our formalism (see previous section).
In this context, and in the absence of an accurate formulation, we estimate the contribution to the
total free-energy arising from the He$^+$, He$^{2+},e^-$ long range interaction
in the framework of the ion-sphere model \cite{salpeter61},
thus considering only the electrostatic contribution to the free-energy.
In this very simplified model,
the interaction between He$^+$ and He$^{2+}$ gives a contribution equal to $Z_1 Z_2 e^2/a$ per pair,
with $Z_1=1$ and $Z_2=2$, whereas the contribution due to the interaction between the central
ion He$^+$ and the uniformly charged sphere $-Z_1e$ gives a contribution
$-3/2 (Z_1e)^2/a$ per He$^+$. 
The He$^{2+}$-e$^-$ and e$^-$-e$^-$ contributions are already included
in the FIP model mentioned in Sec. \ref{sec_fully}. The contribution thus reads:
\begin{equation}
F(N_{\mathrm{He}^+},N_{\mathrm{He}^{2+}})=\frac{N_{\mathrm{He}^+}N_{\mathrm{He}^{2+}}}{2}Z_1 Z_2 {e^2\over a}\,-\,{3\over 2}N_{\mathrm{He}^+}{ (Z_1e)^2\over a}
\label{coulomb}
.
\end{equation}

The very crude treatment of this interaction between He$^+$, He$^{2+}$ and electrons
is certainly a major shortcomings of the present model and Eq. \ref{coulomb} gives at best an
order of magnitude of the contribution of this interaction to the free-energy.
As mentioned above, there is no satisfactory description of ions with bound-states,
He$^+$ in the present context, immersed in a surrounding dense plasma. Indeed, it is difficult
to capture the drastically different nature of the short-range and long-range interactions of such species with surrounding charged particles. This is undoubtedly
a limitation of the chemical picture, and of the related distinction
between different entities. In reality, the concept of identifying He$^+$ or He$^{2+}$ particles,
based on a concept of potential or pseudopotential, becomes
meaningless at high density. Only at high temperature, when kinetic contributions dominate,
is the approach conceptually correct. Therefore, although He$^+$ or He$^{2+}$ are distinguishable
in our model free-energy, we do not pretend to give an accurate description of the second stage
of helium pressure ionization, from He$^{+}$ to He$^{2+}$.
As detailed in the next section, however, we have checked that the present, crude description of the He$^+$-He$^{2+}$ interactions does not alter the final phase diagram. The reason is that, at least in the present model,
 helium pressure ionization proceeds
directly from atomic helium He to fully ionized helium He$^{2+}$, or at least to a strongly ionized stage.
It will certainly be interesting to compare these results with experiments and with results obtained with first-principle calculations, although these latter will certainly have to face their
own difficulties in this complex regime. 

\subsubsection{Summary}
Summarising out the various contributions described in the previous sections,
and following Eq. \ref{eq_HTA} and \ref{eq_FconfHM},
the full model free-energy reads:
\begin{widetext}
\begin{eqnarray}
& &\frac{\beta F}{N_{\mathrm{tot}}}(V,T,\{N_i\})=-\sum_{i={\mathrm{He},\mathrm{He}^{+}}} \frac{N_i}{N_{\mathrm{tot}}}
\left[1+\ln \left(\frac{V}{N_i}\left(\frac{2\pi Mk_BT}{h^2}\right)^{3/2}\right) \right]
-\sum_{i={\mathrm{He},\mathrm{He}^{+}}} \frac{N_i}{N_{\mathrm{tot}}} \ln \sum_{\alpha}
g_{i\alpha} \omega_{i\alpha}^{\mathrm{HS}}\omega_{i\alpha}^{\mu \mathrm{E}} e^{-\beta E_{i\alpha}} \nonumber \\
& &\;\;\;\; +\frac{\beta
F_{\mathrm{HS}}(\{N_{\mathrm{He},\alpha},N_{\mathrm{He}^+,\alpha},
N_{\mathrm{He}^{2+}}\},
\{\sigma_{\mathrm{He},\alpha},\sigma_{\mathrm{He}^+,\alpha},
\sigma_{\mathrm{He}^{2+}}=0\},V,T)
}
{N_{\mathrm{tot}}} \nonumber \\
& &\;\;\;\;
-\sum_{i={\{{\mathrm{He},\alpha}\},\{{\mathrm{He}^+,\alpha}\}}}
\frac{N_i}{N_{\mathrm{tot}}} \frac{\beta \partial F_{\mathrm{HS}}(\{N_i\},
\{\sigma_{\mathrm{He},\alpha},\sigma_{\mathrm{He}^+,\alpha}\},V,T)}
{\partial N_i} 
+ \frac{\beta F_{\mathrm{pol}}(\mathrm{He},\mathrm{He}^+,
\mathrm{He}^{2+},N_{e^-},V,T)}{N_{\mathrm{tot}}}
\nonumber \\
& &\;\;\;\; +\frac{\beta N_{\mathrm{tot}}}{2V}\sum_{i,j=\mathrm{He},\mathrm{He}^{+}}
\frac{N_i}{N_{\mathrm{tot}}}\frac{N_j}{N_{\mathrm{tot}}}\pintun{\bm r} \Phi_{\mathrm{pert}}^{ij}(r)
e^{-\beta \Phi_{\mathrm{ref}}^{ij}(r)}y_{\mathrm{HS}}^{ij}(r) \nonumber \\
& &\;\;\;\; +\frac{\hbar^2}{24(k_BT)^2M_{\mathrm{He}}}\frac{N_{\mathrm{tot}}}{V}
\sum_{i,j=\mathrm{He},\mathrm{He}^{+}}
\frac{N_i}{N_{\mathrm{tot}}}\frac{N_j}{N_{\mathrm{tot}}}\pintun{\bm r} {\bm \nabla}^2\Phi(r)
e^{-\beta \Phi^{ij}(r)}y_{\mathrm{HS}}^{ij}(r) \nonumber \\
& &\;\;\;\; 
+\frac{N_{\mathrm{He}^+}N_{\mathrm{He}^{2+}}}{2 N_{\mathrm{tot}}^2}
Z_1 Z_2 \frac{e^2}{a}
-\frac{3}{2}\frac{N_{\mathrm{He}^+}}{N_{\mathrm{tot}}} \frac{(Z_1e)^2}{a}
+\frac{\beta F^{\mathrm{FIP}}(V,T,N_{\mathrm{He}^{2+}},N_{e^-})}{N_{\mathrm{tot}}}
\label{eq_Fmodelfinal}
,
\end{eqnarray}
\end{widetext}

\noindent where $N_{\mathrm{tot}}=N_{\mathrm{He}}+N_{\mathrm{He}^+}+N_{\mathrm{He}^{2+}}$. Note that $\omega_{i\alpha}^{\mathrm{HS}}
\omega_{i\alpha}^{\mu \mathrm{E}}$
include the occupation probabilities calculated from interactions with neutral
surrounding particles (hard-sphere interaction, Eq. \ref{eq_omegaHS}) and with charged surrounding
particles (microfield interaction, Eq. \ref{eq_omegamicroE}).

The equilibrium populations are derived from the minimization of the free-energy $F(V,T,\{N_i\})$
with respect to two independent variables, given the conditions of
mass conservation, $N_{\mathrm{He}^{2+}}=N_{\mathrm{tot}}
-N_{\mathrm{He}^{+}}-N_{\mathrm{He}}$, and electroneutrality,
$N_{e^-}=N_{\mathrm{He}^{+}}+2N_{\mathrm{He}^{2+}}$:
\begin{equation}
\frac{\partial \frac{\beta F}{N_{\mathrm{tot}}}}{\partial N_{\mathrm{He}}}
\Bigg|_{T,V,N_{\mathrm{He}^+}}=0=
\frac{\partial \frac{\beta F}{N_{\mathrm{tot}}}}{\partial N_{\mathrm{He}^+}}
\Bigg|_{T,V,N_{\mathrm{He}}}
.
\end{equation}
Convergence of this two-dimensional minimization is achieved when the
change in the populations from one iteration to the next one is less than one part
in $3\; 10^{-7}$.
The various
thermodynamic quantities are then calculated from appropriate derivations
of the free-energy.

\section{\label{sec_results} Results}
\begin{figure}
\includegraphics[width=\columnwidth]{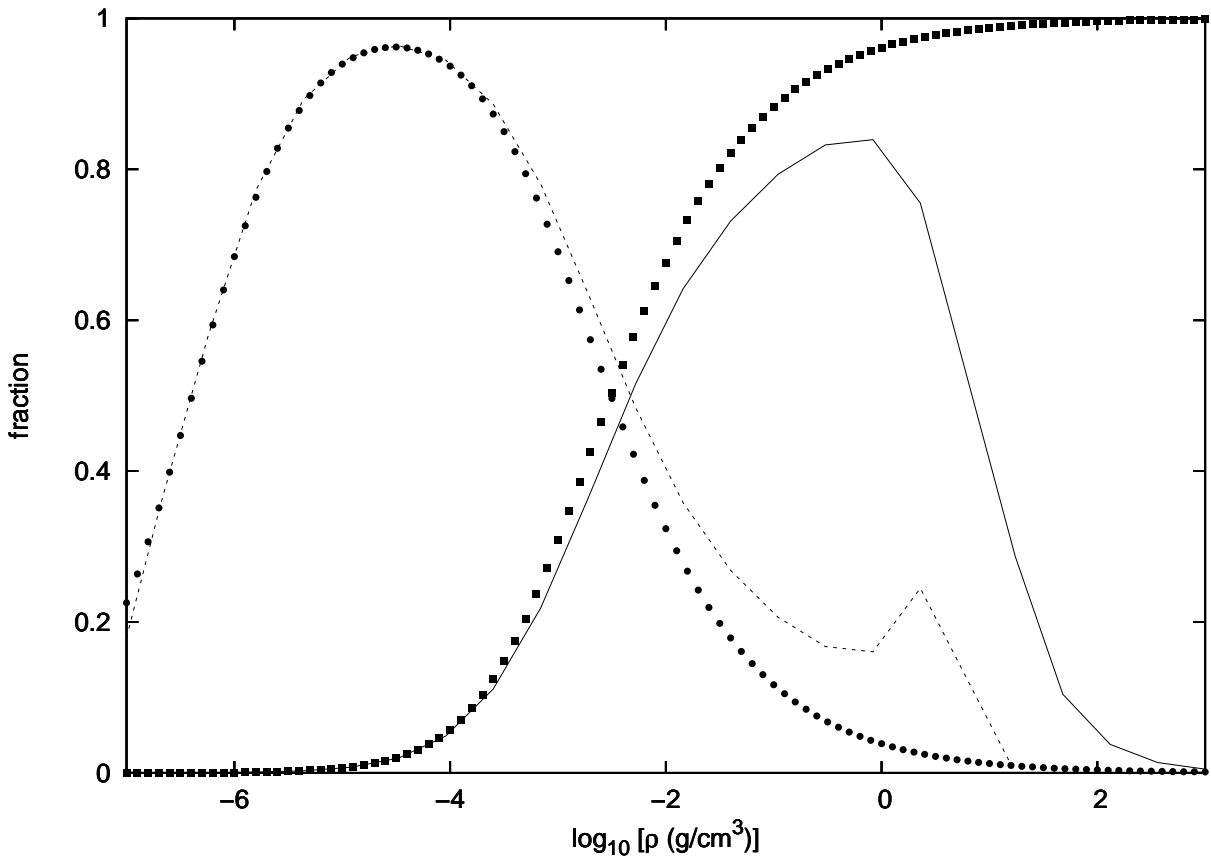}
\caption{\label{fig_saha2} Comparison between the populations obtained with our
model (lines) and those corresponding to the Saha equations (symbols). 
The solid (resp. dotted) line corresponds to the He (resp. He$^+$)
fraction.
The temperature is $T=10^{4.7}$ K.}
\end{figure}

\begin{figure*}
\includegraphics[width=\linewidth]{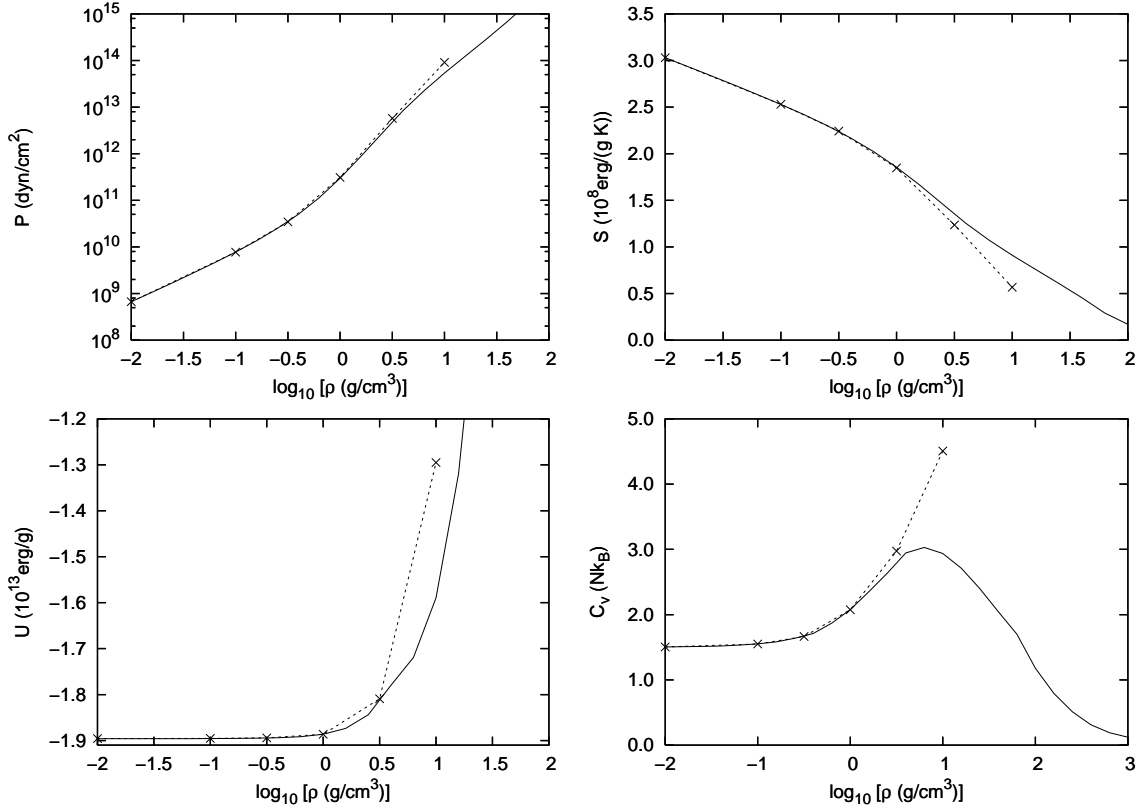}
\caption{\label{fig_ap2} Comparison between the present model (solid lines)
and the results of Aparicio \& Chabrier \cite{aparicio94} (dotted lines
with symbols), which do not include pressure ionization,
for the pressure, the massic entropy, 
the massic internal energy and the specific heat at T=10$^{3.5}$ K.}
\end{figure*}

As mentioned previously, our free-energy model, with the He-He potential
calibrated on sound velocity measurements
\cite{letoullec89}, reproduces the available Hugoniot experiments
\cite{nellis84} (see Fig. \ref{fig_shock}).
We have also checked that we recover the results of the Saha equations in the low-density limit
and, by construction, the fully ionized plasma model at high density.
An example is shown in Fig. \ref{fig_saha2} for $T=10^{4.7}$ K.
The vanishing fraction of bound species populations for $\rho \simgr 10$ g cm$^{-3}$
illustrates the onset of pressure ionization. \\
We have also checked that we recover the results of Aparicio \& Chabrier \cite{aparicio94}
for pure atomic helium in the low-density, low-temperature regime until pressure
ionization sets in (see Fig. \ref{fig_ap2} for T=10$^{3.5}$ K).

\subsection{Limitations of the model} 
As mentioned earlier, our free-energy minimization method is rooted
in the chemical approach. It is based on a heuristic treatment of
the dominant physical effects responsible for the thermodynamic properties of
dense atomic or ionized helium. 
Although it certainly retains some degree of
reality, this model cannot pretend giving an exact description of these
properties, and the results should depend to some extent on the main
approximations used in the model. 
We examine this issue in the present section.

\subsubsection{Lower boundary for $\sigma_1$}
\begin{figure*}
\includegraphics[width=\linewidth]{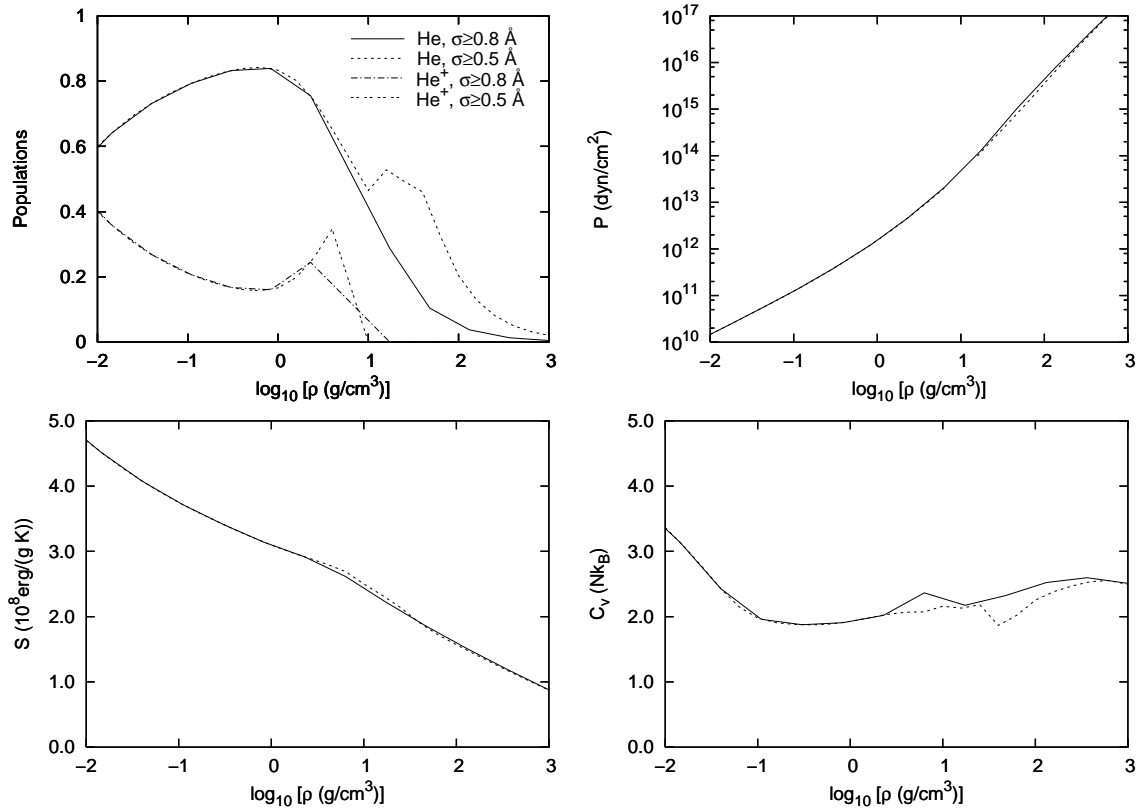}
\caption{\label{fig_troncsigma47} Effect of the lower boundary for
$\sigma_1$ on the populations, the pressure, the massic entropy and
the specific heat as a function of the density. The solid lines correspond
to the case $\sigma_1 \geq 0.8$ \AA, the dotted lines to the
case $\sigma_1 \geq 0.5$ \AA; the temperature is $T=10^{4.7}$ K.}
\end{figure*}

At very high density, the WCA radii tend eventually to zero, as 
shown in Fig. \ref{fig_sigma}. This favors the He and He$^+$ species and thus
prevent pressure ionization to occur, a well identified artifact of the chemical
picture \cite{saumon92} \cite{potekhin96}. In order
to prevent such an unphysical behaviour, we define arbitrarily a lower limit
for $\sigma_{\mathrm{He}}$ and $\sigma_{\mathrm{He}^+}$. 
Fig. \ref{fig_troncsigma47} (for the 10$^{4.7}$ K isotherm) illustrates
the effect of this 
approximation
for $\sigma_1 \geq 0.8$ \AA~and $\sigma_1 \geq 0.5$ \AA.

Not surprisingly, the choice of a lower limit for $\sigma_1$ affects appreciably
the populations in the very regime of pressure ionization. However the effect is almost inconsequential on the 
thermodynamic quantities, the very purpose of the present calculations.
This stems from the fact that the bound-species do not
contribute to the free-energy when they are associated with a very small radius.
The final model calculations were made with $\sigma_1 = 0.8$ \AA.

\subsubsection{\label{sec_polariz} Polarizability of He-He$^+$}
We have also tested the influence of the polarizability $\alpha_i$
which appears in the He-He$^+$ potential, and which has been taken
equal to the volume $\sigma_{\mathrm{He-He^+}}^3$. Calculations conducted
with a value of $\alpha_i$ reduced or increased by a factor 10 left
the results nearly unaffected. This can be easily 
understood as in the domain where non-ideal effects play a role, 
He and He$^+$ do not coexist in comparable fractions most of the time.
Moreover, the contribution of $F_{\mathrm{pol}}$ to the total free-energy remains always
marginal. 

\subsubsection{Validity of the quantum correction $F_{\mathrm{qm}}$}
As mentioned in Sec. \ref{sec_Fqm}, we have used the first-order term of the
Wigner-Kirkwood expansion to take into account the quantum effects between
atomic centers. This expansion becomes invalid at high density and low temperature.
As a rule of thumb, the domain of reliability of the expansion is given by:
$\log_{10}T_{\mathrm{K}}-\log_{10}
\rho _{{\mathrm{g/cm}}^3}
\simgr 2$. Such a limitation has no consequence in an astrophysical
context, as no astrophysical object with a helium composition exists beyond this limit.

\subsubsection{\label{sec_testHe1He2coupling} Influence of the 
He$^+$-He$^{2+}$ coupling}
\begin{figure*}
\includegraphics[width=\linewidth]{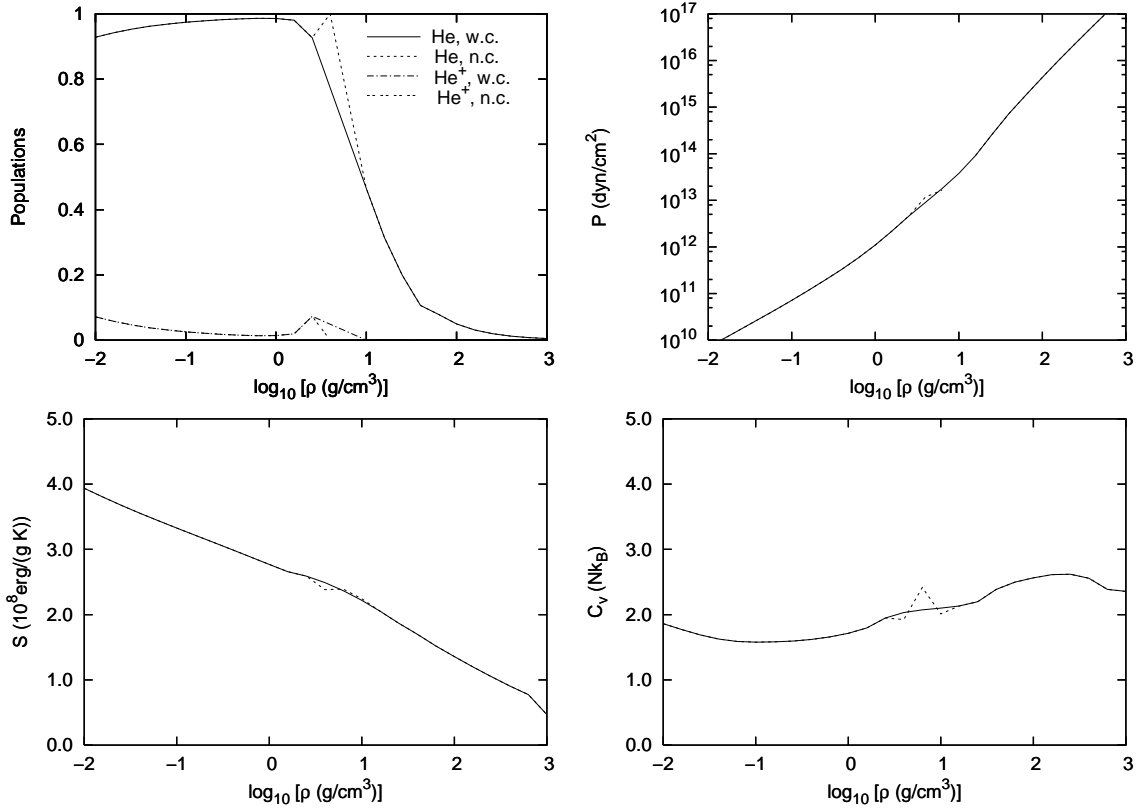}
\caption{\label{fig_test_coupling45}{Effect of the
description of the He$^+$-He$^{2+}$ and He$^+$-e$^-$ couplings on the
populations (w.c $\equiv$ with coupling, n.c $\equiv$ no coupling), the
pressure, the massic entropy and the specific heat as a function of the
density. If not specified, the solid line corresponds to
the case with coupling and the dotted line to the case without any coupling;
the temperature is $T=10^{4.5}$ K.}}
\end{figure*}

As mentioned in Sec. \ref{sec_FHe1He2}, the long range interaction between He$^+$, He$^{2+}$ and e$^-$
is treated in a rather crude way in the present model. 
We have tested the influence of this approximation by submitting a few tests without this term. The results are illustrated
in Fig. \ref{fig_test_coupling45} for $T=10^{4.5}$ K. 
We have
checked other isotherms, and the conclusion is that the EOS and its derivatives
are nearly independent of this coupling term except in a very limited
temperature-density range. 
As illustrated in the next section, but also on Fig. \ref{fig_saha2}, 
\ref{fig_troncsigma47} and \ref{fig_test_coupling45}, 
the reason is that pressure ionization
occurs directly from He to He$^{2+}$, with no regions where He$^{+}$ and He$^{2+}$ coexist in comparable number, 
except at high temperature ($T\simgr 10^{5}$ K) where temperature ionization dominates.
Although we certainly cannot rule out the fact that this is an artefact of our
model,
a possible physical explanation might be the large
differences between the ground-state energies of the different species, much
larger than
for hydrogen. The contribution of the ground-state energy of He to the total
free-energy thus prevents partial ionization to occur, favoring the atomic
phase.
As mentioned previously, it will be interesting to compare this prediction with experimental results
and first-principle
calculations, once they will be available, to verify whether this behaviour is a flaw of the present
model or whether it reflects the behaviour of helium pressure ionization, an extremely
interesting issue.

\subsection{Thermodynamical quantities }
\begin{figure}
\includegraphics[width=\columnwidth]{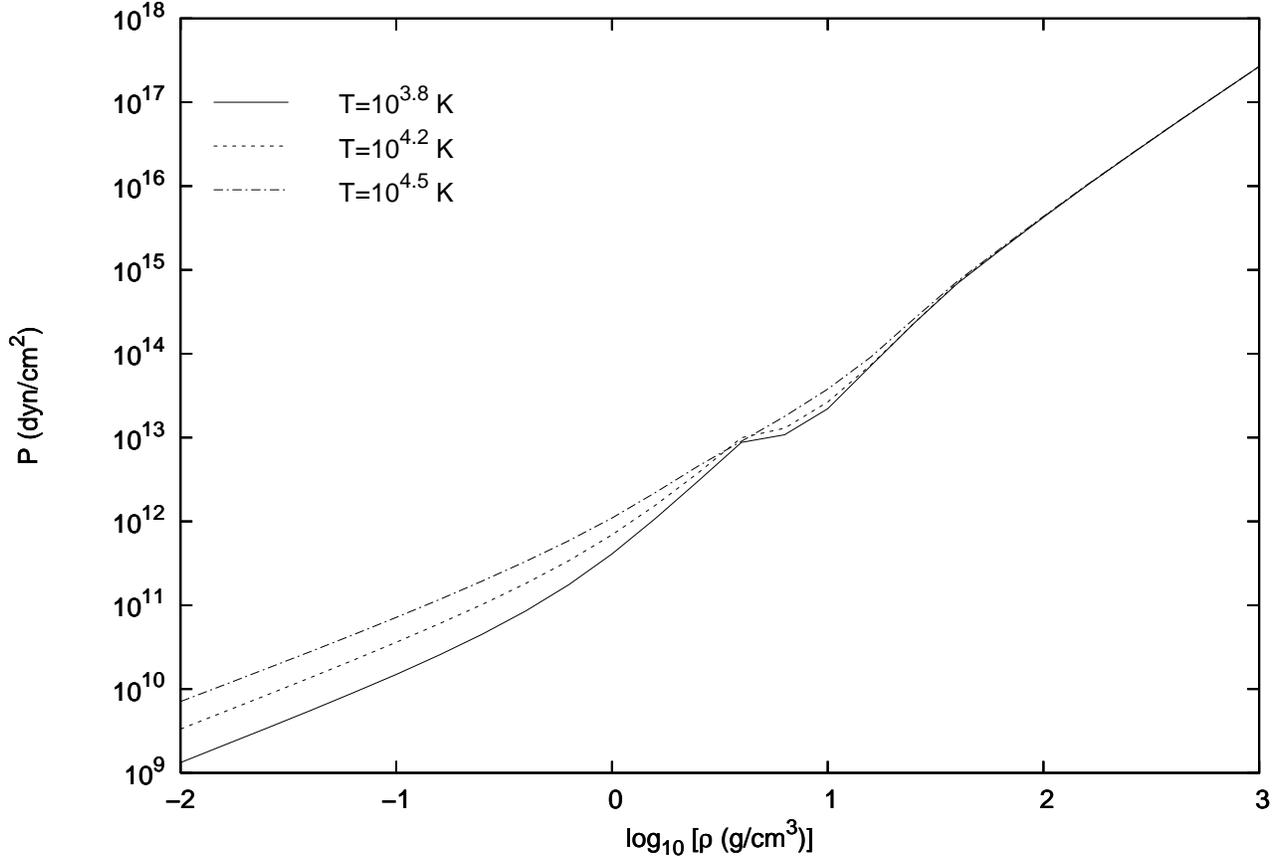}
\caption{\label{fig_res_P} Pressure as a function of the density for three
isotherms. Solid line: $T=10^{3.8}$ K, dotted-line: $T=10^{4.2}$ K,
dot-dashed line: $T=10^{4.5}$ K.}
\end{figure}

\begin{figure}
\includegraphics[width=\columnwidth]{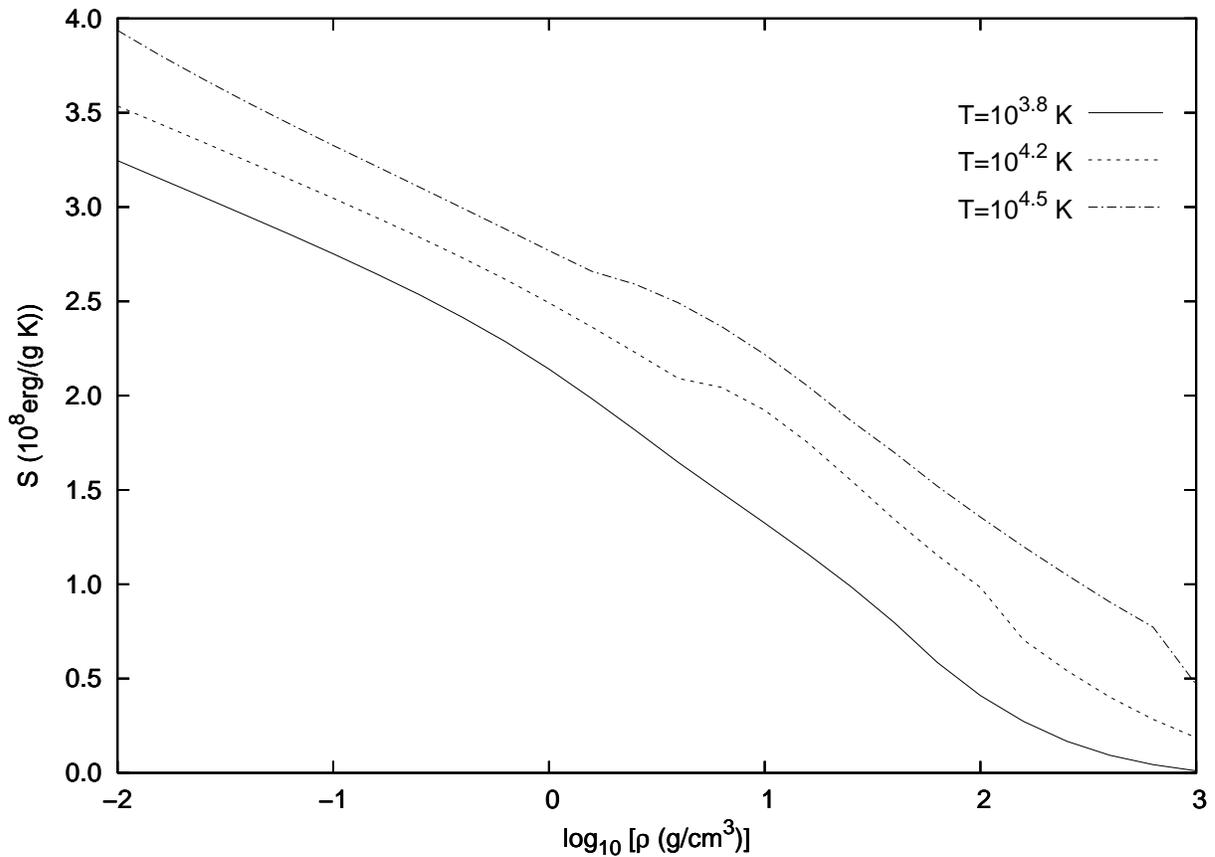}
\caption{\label{fig_res_S} Same as Fig. \ref{fig_res_P} for the massic entropy.}
\end{figure}

\begin{figure}
\includegraphics[width=\columnwidth]{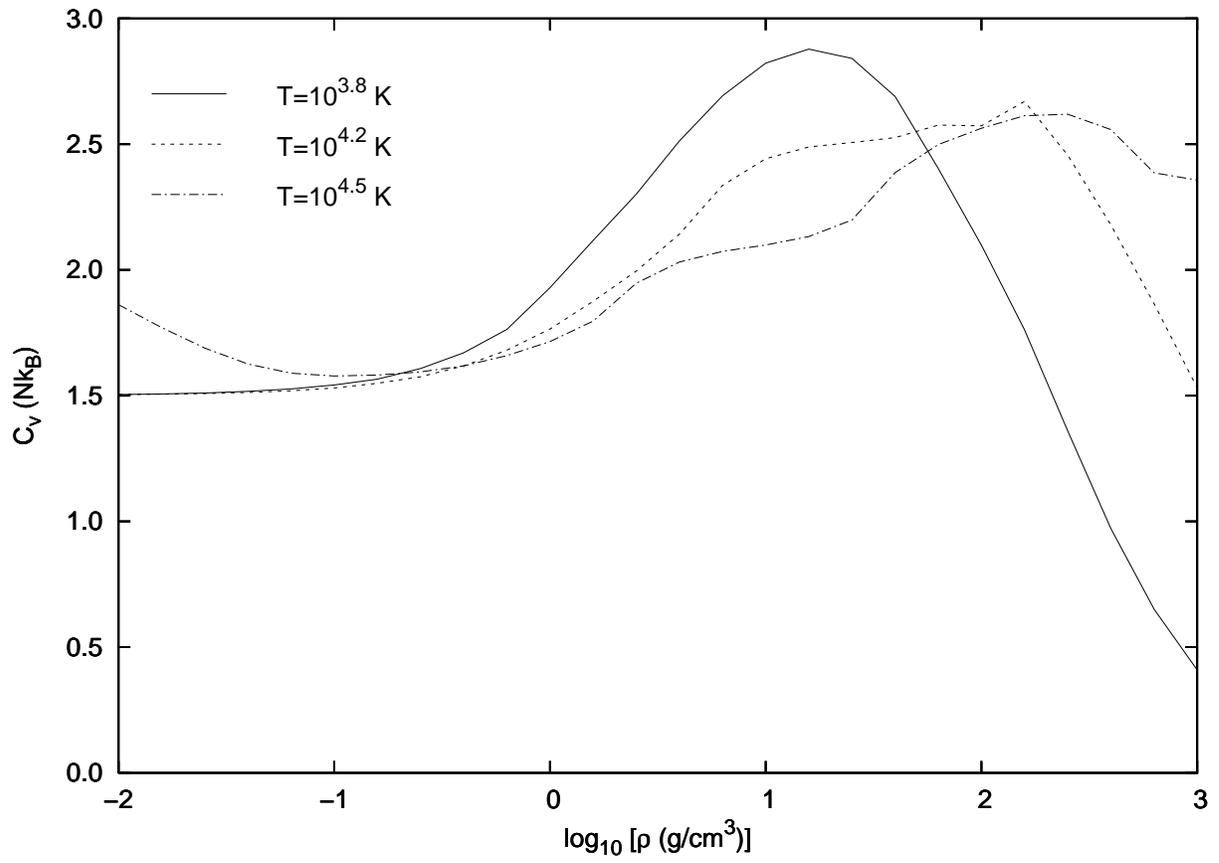}
\caption{\label{fig_res_Cv}  Same as Fig. \ref{fig_res_P} for the specific heat.}
\end{figure}

\begin{figure}
\includegraphics[width=\columnwidth]{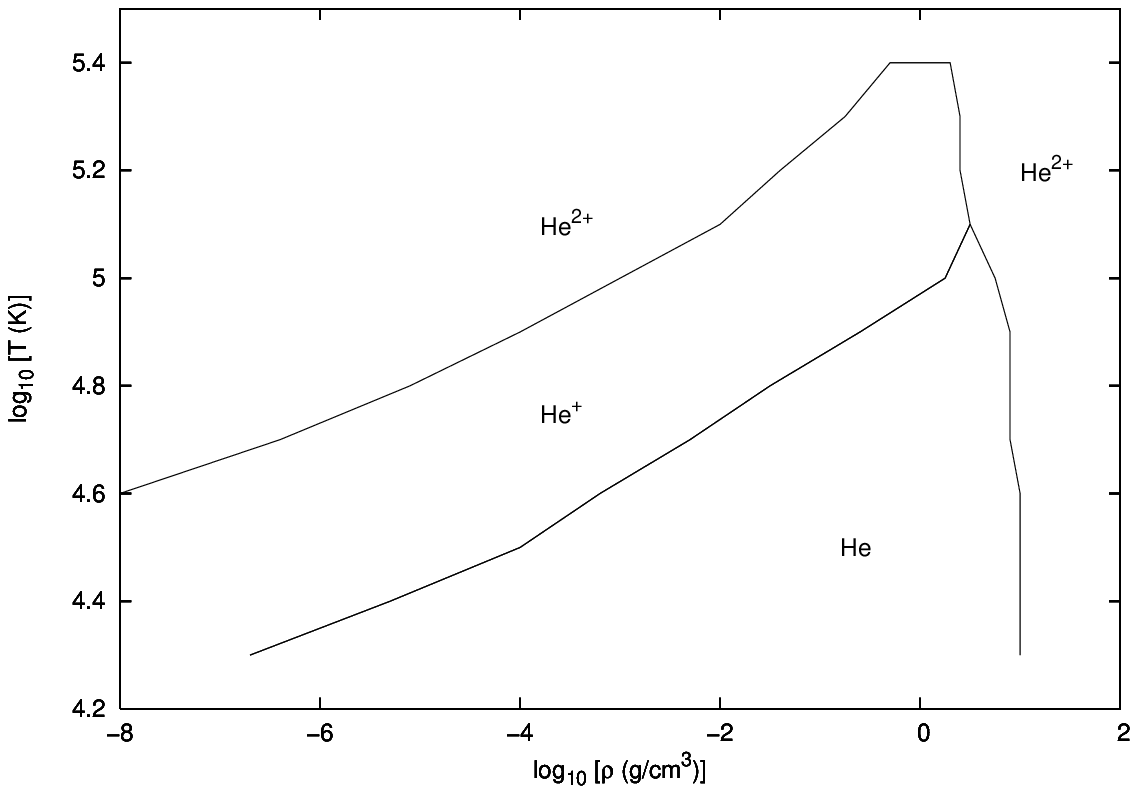}
\caption{\label{fig_diagpredom} Phase diagram of helium. The lines
separate the different domains of predominance of the different
species He, He$^+$ or He$^{2+}$.}
\end{figure}

A subset of our final EOS calculations, based on the model free-energy 
(\ref{eq_Fmodelfinal})
is presented in Tables \ref{tab_T38}-\ref{tab_T45},
corresponding to Fig. \ref{fig_res_P} (for the pressure), 
Fig. \ref{fig_res_S} (for the massic entropy) and Fig. \ref{fig_res_Cv} (for
the specific heat).
For these calculations, ten internal levels have been considered, 
both for He and He$^+$. These ten levels are enough to represent the 
internal partition function as the highest levels are always destroyed
even for the lowest density we are considering. 
No He doubly excited states have been considered in our calculation.
This is a reasonable approximation because of the two following reasons.
First of all, the high energy cost of these states (the first
doubly excited state lies $\sim 60$~eV above the He ground state)
disfavors their formation (in a way similar to the direct ionization of
He to He$^{2+}$ without any He$^{+}$ state, see following discussion).
The second reason is their rapid
decay by autoionization (typically in $10^{-13}-10^{-14}$~s).
It is therefore unlikely that these states survive in the midst of
interacting neighbour particles.
The zero of energy 
corresponds to the fully ionized plasma at zero temperature.
The rising behaviour of $C_V$ for $\log_{10} \rho _{{\mathrm{g/cm}}^3} \simgr -1$ stems from correlations between
helium atoms (configurational free-energy), since all excited levels are destroyed at this density,
at least for the coolest temperature. The drop at larger density reflects pressure ionization, from
He to He$^{2+}$.
We also present in
Fig. \ref{fig_diagpredom} the phase diagram of helium.
The lines
separate the different domains where either He, He$^+$ or He$^{2+}$ is
the dominant species, {\it i.e.} represents a fraction larger than 50 \%. 
As mentioned previously, an interesting prediction of this diagram (see also the Tables) is that for
$T\simlw 10^5$ K, pressure ionization, defined as $x_{\mathrm{He}^{2+}}\simgr 0.5$, 
proceeds
directly from He to He$^{2+}$ at $\rho \simgr$ 10  g cm$^{-3}$, {\it i.e.}
$P \simgr 20$ Mbar. 
As mentioned in Sec. \ref{sec_testHe1He2coupling},
the sharp transition due to pressure ionization, from $x_{\mathrm{He}}\simgr 0.5$ to $x_{\mathrm{He}^{2+}}\simgr 0.5$ at $\rho \sim$ 10  g cm$^{-3}$ (see Tables), and
the persistence of atomic helium at high density, might reflect the large
energy cost of the ground-state energies of ionized species 
(24.587 eV and 79.003 eV) to the total
free-energy. Eventually,
abrupt ionization occurs
from He to He$^{2+}$,
unless temperature is high enough
to unbound one of the two electrons from the helium atom.
This is corroborated by
the fact that the pressure ionization of He$^+$ (which happens if 
$T\simgr 10^{5}$ K) occurs at lower densities, $\rho \simgr 1$ g cm$^{-3}$.
This phase diagram can be compared with the one for hydrogen
\cite{saumon95}. For deuterium,
the EOS is essentially
the same as for hydrogen providing the nucleus mass is rescaled
\cite{saumon00}. However, for helium,
because of the $Z=2$ nucleus and the induced electronic structures, the
phase diagram is different, and pressure
ionization occurs at larger pressures than for H or D.

\begin{table*}
\caption{Equation of state for the isotherm $T=10^{3.8}$ K. For each value 
of the density are given the abundances of He, He$^+$ and He$^{2+}$, the 
pressure, the massic entropy, the massic internal energy (with a zero of
energy corresponding to the fully ionized plasma at zero temperature) and the
specific heat.}
\label{tab_T38}
\begin{ruledtabular}
\begin{tabular}{cccccccc}
$\log_{10}(\rho/ 1$ g/cm$^3)$ & $x_{\mathrm{He}}$ & $x_{\mathrm{He}^+}$ & $x_{\mathrm{He}^{2+}}$ &
      $P$ (dyn/cm$^2$)          & $S$ (erg/g/K)       & $U$ (erg/g)           & $C_v (Nk_B)$\\
  $-2.00$ & $1.0000 \times 10^{0} $ & $0.0000 \times 10^{0} $ & $0.0000 \times 10^{0}$ & $0.1330 \times 10^{10}$ & $0.3246 \times 10^{9}$ & $-0.1886 \times 10^{14}$ & $0.1504 \times 10^{1}$ \\
  $-1.60$ & $1.0000 \times 10^{0} $ & $0.0000 \times 10^{0} $ & $0.0000 \times 10^{0}$ & $0.3409 \times 10^{10}$ & $0.3052 \times 10^{9}$ & $-0.1886 \times 10^{14}$ & $0.1510 \times 10^{1}$ \\
  $-1.20$ & $1.0000 \times 10^{0} $ & $0.0000 \times 10^{0} $ & $0.0000 \times 10^{0}$ & $0.9003 \times 10^{10}$ & $0.2854 \times 10^{9}$ & $-0.1886 \times 10^{14}$ & $0.1527 \times 10^{1}$ \\
  $-0.80$ & $1.0000 \times 10^{0} $ & $0.0000 \times 10^{0} $ & $0.0000 \times 10^{0}$ & $0.2554 \times 10^{11}$ & $0.2646 \times 10^{9}$ & $-0.1885 \times 10^{14}$ & $0.1565 \times 10^{1}$ \\
  $-0.40$ & $1.0000 \times 10^{0} $ & $0.0000 \times 10^{0} $ & $0.0000 \times 10^{0}$ & $0.8538 \times 10^{11}$ & $0.2415 \times 10^{9}$ & $-0.1883 \times 10^{14}$ & $0.1670 \times 10^{1}$ \\
  $ 0.00$ & $1.0000 \times 10^{0} $ & $0.0000 \times 10^{0} $ & $0.0000 \times 10^{0}$ & $0.4092 \times 10^{12}$ & $0.2141 \times 10^{9}$ & $-0.1873 \times 10^{14}$ & $0.1929 \times 10^{1}$ \\
  $ 0.40$ & $1.0000 \times 10^{0} $ & $0.0000 \times 10^{0} $ & $0.0000 \times 10^{0}$ & $0.3037 \times 10^{13}$ & $0.1816 \times 10^{9}$ & $-0.1828 \times 10^{14}$ & $0.2301 \times 10^{1}$ \\
  $ 0.60$ & $1.0000 \times 10^{0} $ & $0.0000 \times 10^{0} $ & $0.0000 \times 10^{0}$ & $0.8729 \times 10^{13}$ & $0.1646 \times 10^{9}$ & $-0.1762 \times 10^{14}$ & $0.2513 \times 10^{1}$ \\
  $ 0.80$ & $0.7158 \times 10^{0} $ & $0.1688 \times 10^{-4}$ & $0.2842 \times 10^{0}$ & $0.1082 \times 10^{14}$ & $0.1484 \times 10^{9}$ & $-0.1632 \times 10^{14}$ & $0.2693 \times 10^{1}$ \\
  $ 1.00$ & $0.4316 \times 10^{0} $ & $0.2531 \times 10^{-4}$ & $0.5684 \times 10^{0}$ & $0.2210 \times 10^{14}$ & $0.1324 \times 10^{9}$ & $-0.1502 \times 10^{14}$ & $0.2822 \times 10^{1}$ \\
  $ 1.20$ & $0.3144 \times 10^{0} $ & $0.2531 \times 10^{-4}$ & $0.6856 \times 10^{0}$ & $0.7143 \times 10^{14}$ & $0.1161 \times 10^{9}$ & $-0.1307 \times 10^{14}$ & $0.2878 \times 10^{1}$ \\
  $ 1.40$ & $0.1983 \times 10^{0} $ & $0.2531 \times 10^{-4}$ & $0.8017 \times 10^{0}$ & $0.2314 \times 10^{15}$ & $0.9875 \times 10^{8}$ & $-0.8728 \times 10^{13}$ & $0.2841 \times 10^{1}$ \\
  $ 1.60$ & $0.1062 \times 10^{0} $ & $0.1688 \times 10^{-4}$ & $0.8938 \times 10^{0}$ & $0.6832 \times 10^{15}$ & $0.7975 \times 10^{8}$ & $-0.5620 \times 10^{12}$ & $0.2689 \times 10^{1}$ \\
  $ 2.00$ & $0.4980 \times 10^{-1}$ & $0.1688 \times 10^{-4}$ & $0.9502 \times 10^{0}$ & $0.4249 \times 10^{16}$ & $0.4101 \times 10^{8}$ & $ 0.2846 \times 10^{14}$ & $0.2098 \times 10^{1}$ \\
  $ 2.40$ & $0.1980 \times 10^{-1}$ & $0.1688 \times 10^{-4}$ & $0.9802 \times 10^{0}$ & $0.2338 \times 10^{17}$ & $0.1681 \times 10^{8}$ & $ 0.9215 \times 10^{14}$ & $0.1360 \times 10^{1}$ \\
  $ 2.80$ & $0.7880 \times 10^{-2}$ & $0.1125 \times 10^{-4}$ & $0.9921 \times 10^{0}$ & $0.1197 \times 10^{18}$ & $0.4317 \times 10^{7}$ & $ 0.2210 \times 10^{15}$ & $0.6508 \times 10^{0}$
\end{tabular}
\end{ruledtabular}
\end{table*}

\begin{table*}
\caption{Equation of state for the isotherm $T=10^{4.2}$ K. For each value
of the density are given the abundances of He, He$^+$ and He$^{2+}$, the
pressure, the massic entropy, the massic internal energy (with a zero of
energy corresponding to the fully ionized plasma at zero temperature) and the
specific heat.}
\label{tab_T42}
\begin{ruledtabular}
\begin{tabular}{cccccccc}
$\log_{10}(\rho/ 1$ g/cm$^3)$ & $x_{\mathrm{He}}$ & $x_{\mathrm{He}^+}$ & $x_{\mathrm{He}^{2+}}$ &
      $P$ (dyn/cm$^2$)          & $S$ (erg/g/K)       & $U$ (erg/g)           & $C_v (Nk_B)$\\
 $-2.00$ & $0.9995 \times 10^{0}$  & $0.4950 \times 10^{-3}$ & $0.0000 \times 10^{0}$ & $0.3327 \times 10^{10}$ & $0.3536 \times 10^{9}$ & $-0.1856 \times 10^{14}$ & $0.1505 \times 10^{1}$ \\
 $-1.60$ & $0.9997 \times 10^{0}$  & $0.3000 \times 10^{-3}$ & $0.0000 \times 10^{0}$ & $0.8474 \times 10^{10}$ & $0.3342 \times 10^{9}$ & $-0.1856 \times 10^{14}$ & $0.1508 \times 10^{1}$ \\
 $-1.20$ & $0.9998 \times 10^{0}$  & $0.1800 \times 10^{-3}$ & $0.0000 \times 10^{0}$ & $0.2203 \times 10^{11}$ & $0.3146 \times 10^{9}$ & $-0.1856 \times 10^{14}$ & $0.1519 \times 10^{1}$ \\
 $-0.80$ & $0.9999 \times 10^{0}$  & $0.1100 \times 10^{-3}$ & $0.0000 \times 10^{0}$ & $0.6004 \times 10^{11}$ & $0.2944 \times 10^{9}$ & $-0.1854 \times 10^{14}$ & $0.1549 \times 10^{1}$ \\
 $-0.40$ & $0.9999 \times 10^{0}$  & $0.6750 \times 10^{-4}$ & $0.0000 \times 10^{0}$ & $0.1817 \times 10^{12}$ & $0.2730 \times 10^{9}$ & $-0.1850 \times 10^{14}$ & $0.1618 \times 10^{1}$ \\
 $ 0.00$ & $0.9999 \times 10^{0}$  & $0.6000 \times 10^{-4}$ & $0.0000 \times 10^{0}$ & $0.6898 \times 10^{12}$ & $0.2493 \times 10^{9}$ & $-0.1837 \times 10^{14}$ & $0.1765 \times 10^{1}$ \\
 $ 0.40$ & $1.0000 \times 10^{0}$  & $0.0000 \times 10^{0} $ & $0.0000 \times 10^{0}$ & $0.3814 \times 10^{13}$ & $0.2228 \times 10^{9}$ & $-0.1785 \times 10^{14}$ & $0.1996 \times 10^{1}$ \\
 $ 0.60$ & $1.0000 \times 10^{0}$  & $0.0000 \times 10^{0} $ & $0.0000 \times 10^{0}$ & $0.9964 \times 10^{13}$ & $0.2091 \times 10^{9}$ & $-0.1717 \times 10^{14}$ & $0.2142 \times 10^{1}$ \\
 $ 0.80$ & $0.7325 \times 10^{0}$  & $0.1688 \times 10^{-4}$ & $0.2675 \times 10^{0}$ & $0.1295 \times 10^{14}$ & $0.2043 \times 10^{9}$ & $-0.1593 \times 10^{14}$ & $0.2336 \times 10^{1}$ \\
 $ 1.00$ & $0.4650 \times 10^{0}$  & $0.2531 \times 10^{-4}$ & $0.5350 \times 10^{0}$ & $0.2660 \times 10^{14}$ & $0.1923 \times 10^{9}$ & $-0.1469 \times 10^{14}$ & $0.2442 \times 10^{1}$ \\
 $ 1.20$ & $0.2405 \times 10^{0}$  & $0.2531 \times 10^{-4}$ & $0.7595 \times 10^{0}$ & $0.7335 \times 10^{14}$ & $0.1751 \times 10^{9}$ & $-0.1257 \times 10^{14}$ & $0.2489 \times 10^{1}$ \\
 $ 1.40$ & $0.1983 \times 10^{0}$  & $0.2531 \times 10^{-4}$ & $0.8017 \times 10^{0}$ & $0.2305 \times 10^{15}$ & $0.1551 \times 10^{9}$ & $-0.8125 \times 10^{13}$ & $0.2507 \times 10^{1}$ \\
 $ 1.60$ & $0.1062 \times 10^{0}$  & $0.1688 \times 10^{-4}$ & $0.8938 \times 10^{0}$ & $0.7012 \times 10^{15}$ & $0.1345 \times 10^{9}$ & $ 0.1229 \times 10^{12}$ & $0.2526 \times 10^{1}$ \\
 $ 2.00$ & $0.4980 \times 10^{-1}$ & $0.1688 \times 10^{-4}$ & $0.9502 \times 10^{0}$ & $0.4294 \times 10^{16}$ & $0.9833 \times 10^{8}$ & $ 0.2923 \times 10^{14}$ & $0.2573 \times 10^{1}$ \\
 $ 2.40$ & $0.1980 \times 10^{-1}$ & $0.1688 \times 10^{-4}$ & $0.9802 \times 10^{0}$ & $0.2346 \times 10^{17}$ & $0.5425 \times 10^{8}$ & $ 0.9255 \times 10^{14}$ & $0.2456 \times 10^{1}$ \\
 $ 2.80$ & $0.7880 \times 10^{-2}$ & $0.1125 \times 10^{-4}$ & $0.9921 \times 10^{0}$ & $0.1199 \times 10^{18}$ & $0.2836 \times 10^{8}$ & $ 0.2213 \times 10^{15}$ & $0.1865 \times 10^{1}$
\end{tabular}
\end{ruledtabular}
\end{table*}

\begin{table*}
\caption{Equation of state for the isotherm $T=10^{4.5}$ K. For each value
of the density are given the abundances of He, He$^+$ and He$^{2+}$, the
pressure, the massic entropy, the massic internal energy (with a zero of
energy corresponding to the fully ionized plasma at zero temperature) and the
specific heat.}
\label{tab_T45}
\begin{ruledtabular}
\begin{tabular}{cccccccc}
$\log_{10}(\rho/ 1$ g/cm$^3)$ & $x_{\mathrm{He}}$ & $x_{\mathrm{He}^+}$ & $x_{\mathrm{He}^{2+}}$ &
      $P$ (dyn/cm$^2$)          & $S$ (erg/g/K)       & $U$ (erg/g)           & $C_v (Nk_B)$\\
  $-2.00$ & $0.9287 \times 10^{0} $ & $0.7128 \times 10^{-1}$ & $0.0000 \times 10^{0}$ & $0.7083 \times 10^{10}$ & $0.3936 \times 10^{9}$ & $-0.1758 \times 10^{14}$ & $0.1862 \times 10^{1}$ \\
  $-1.60$ & $0.9532 \times 10^{0} $ & $0.4680 \times 10^{-1}$ & $0.0000 \times 10^{0}$ & $0.1757 \times 10^{11}$ & $0.3676 \times 10^{9}$ & $-0.1775 \times 10^{14}$ & $0.1689 \times 10^{1}$ \\
  $-1.20$ & $0.9690 \times 10^{0} $ & $0.3102 \times 10^{-1}$ & $0.0000 \times 10^{0}$ & $0.4441 \times 10^{11}$ & $0.3439 \times 10^{9}$ & $-0.1786 \times 10^{14}$ & $0.1590 \times 10^{1}$ \\
  $-0.80$ & $0.9788 \times 10^{0} $ & $0.2124 \times 10^{-1}$ & $0.0000 \times 10^{0}$ & $0.1169 \times 10^{12}$ & $0.3216 \times 10^{9}$ & $-0.1790 \times 10^{14}$ & $0.1582 \times 10^{1}$ \\
  $-0.40$ & $0.9848 \times 10^{0} $ & $0.1524 \times 10^{-1}$ & $0.0000 \times 10^{0}$ & $0.3322 \times 10^{12}$ & $0.2995 \times 10^{9}$ & $-0.1789 \times 10^{14}$ & $0.1619 \times 10^{1}$ \\
  $ 0.00$ & $0.9862 \times 10^{0} $ & $0.1376 \times 10^{-1}$ & $0.0000 \times 10^{0}$ & $0.1098 \times 10^{13}$ & $0.2769 \times 10^{9}$ & $-0.1773 \times 10^{14}$ & $0.1715 \times 10^{1}$ \\
  $ 0.40$ & $0.9271 \times 10^{0} $ & $0.7296 \times 10^{-1}$ & $0.0000 \times 10^{0}$ & $0.4567 \times 10^{13}$ & $0.2590 \times 10^{9}$ & $-0.1699 \times 10^{14}$ & $0.1947 \times 10^{1}$ \\
  $ 0.60$ & $0.8063 \times 10^{0} $ & $0.7988 \times 10^{-1}$ & $0.1139 \times 10^{0}$ & $0.9005 \times 10^{13}$ & $0.2492 \times 10^{9}$ & $-0.1651 \times 10^{14}$ & $0.2031 \times 10^{1}$ \\
  $ 0.80$ & $0.6457 \times 10^{0} $ & $0.5800 \times 10^{-1}$ & $0.2964 \times 10^{0}$ & $0.1788 \times 10^{14}$ & $0.2366 \times 10^{9}$ & $-0.1527 \times 10^{14}$ & $0.2073 \times 10^{1}$ \\
  $ 1.00$ & $0.4726 \times 10^{0} $ & $0.2536 \times 10^{-1}$ & $0.5020 \times 10^{0}$ & $0.3791 \times 10^{14}$ & $0.2218 \times 10^{9}$ & $-0.1403 \times 10^{14}$ & $0.2099 \times 10^{1}$ \\
  $ 1.20$ & $0.3144 \times 10^{0} $ & $0.2531 \times 10^{-4}$ & $0.6856 \times 10^{0}$ & $0.9098 \times 10^{14}$ & $0.2050 \times 10^{9}$ & $-0.1189 \times 10^{14}$ & $0.2132 \times 10^{1}$ \\
  $ 1.40$ & $0.1983 \times 10^{0} $ & $0.2531 \times 10^{-4}$ & $0.8017 \times 10^{0}$ & $0.2619 \times 10^{15}$ & $0.1867 \times 10^{9}$ & $-0.7390 \times 10^{13}$ & $0.2198 \times 10^{1}$ \\
  $ 1.60$ & $0.1062 \times 10^{0} $ & $0.1688 \times 10^{-4}$ & $0.8938 \times 10^{0}$ & $0.7280 \times 10^{15}$ & $0.1700 \times 10^{9}$ & $ 0.9305 \times 10^{12}$ & $0.2387 \times 10^{1}$ \\
  $ 2.00$ & $0.4980 \times 10^{-1}$ & $0.1688 \times 10^{-4}$ & $0.9502 \times 10^{0}$ & $0.4359 \times 10^{16}$ & $0.1356 \times 10^{9}$ & $ 0.3008 \times 10^{14}$ & $0.2563 \times 10^{1}$ \\
  $ 2.40$ & $0.1980 \times 10^{-1}$ & $0.1688 \times 10^{-4}$ & $0.9802 \times 10^{0}$ & $0.2364 \times 10^{17}$ & $0.1049 \times 10^{9}$ & $ 0.9382 \times 10^{14}$ & $0.2619 \times 10^{1}$ \\
  $ 2.80$ & $0.7880 \times 10^{-2}$ & $0.1125 \times 10^{-4}$ & $0.9921 \times 10^{0}$ & $0.1203 \times 10^{18}$ & $0.7731 \times 10^{8}$ & $ 0.2226 \times 10^{15}$ & $0.2386 \times 10^{1}$
\end{tabular}
\end{ruledtabular}
\end{table*}

\section{\label{sec_concl} Conclusion}

In this paper, we have computed a free-energy model aimed at deriving
the thermodynamic quantities of dense fluid helium, from the low-density
atomic domain to the high-density fully ionized regime, covering the regime of
partial ionization. The model is based on the so-called chemical picture for the
description of the interactions between the different species in the fluid. The
abundances of the various atomic and ionic components are obtained through minimization
of the free-energy. Despite the shortcomings inherent
to the chemical approach, we believe the present model to give a reasonable
description of the equation of state of dense helium, including the regime of pressure ionization.
Although the basis of the model become of doubtful validity in this latter domain, this
affects only limited regions of the temperature-density diagram. Comparisons with
available sound speed measurements and shock-wave experiments for atomic helium
assess the validity of the model up to the megabar range, whereas at very high
density the model recovers the fully ionized plasma model and thus Monte-Carlo
simulations of the thermodynamic properties of the so-called one-component plasma
(OCP) model. Although the present model cannot pretend giving a precise determination
of the various atomic and ionic concentrations in the fluid, at least in the pressure ionization regimes,
it yields a reasonably
accurate determination of the phase diagram of dense, fluid helium with its
various He/He$^{+}$/He$^{2+}$ ionization contours. For $T\simlw 10^5$ K, pressure ionization
is found to occur directly from atomic helium He to fully ionized helium 
He$^{2+}$, or at least to a strongly ionized state,
without
He$^{+}$ stage ($x_{\mathrm{He}^+} <$ a few \%). It would be interesting to test such a prediction with high-pressure dynamical
experiments. Indeed, such a behaviour of the phase diagram bears important consequences for
the thermodynamic, magnetic and transport properties of the interior of cool and dense astrophysical
objects, including giant planets.
In all cases, {\it pressure} ionization is found to occur around $\rho \sim$
10 g cm$^{-3}$,
{\it i.e.} $P\sim 20 \times 10^6$ bar. 
Detailed explorations of the sensitivity of the results upon various
approximations
entering the free-energy model show that they remain inconsequential
on the first derivatives of the free-energy over most of the phase diagram.
In some limited regions, however, characteristic of the pressure ionization
regime, maximum variations of the entropy and the pressure can reach $\sim 5\%$
and $\sim 20\%$, respectively, in the worst case.
Although still modest in most cases, the uncertainties become
larger for second derivatives, in particular the ones directly related to the
different
degrees of freedom and thus to the relative populations, like the specific
heat. As
mentioned above, however,
only limited regions of the phase diagram are concerned by the regime
where various species coexist in comparable numbers.
As a whole, the present model remains simple enough to allow the calculation of the
EOS of dense helium over an extended domain of pressure and density, a necessary
condition for applications to the computation of stellar and giant planet internal
structure and high-pressure experiment diagnostics.

Besides its astrophysical interest, the calculation of the phase diagram of dense helium is of intrinsic
theoretical interest. Indeed, comparison betwen these calculations and
near-future high-pressure shock-wave
or laser experiments will allow a better determination of the domains of validity
of the present model and of the possible improvements. By such, these comparisons
will yield a better understanding of the
properties of matter under extreme conditions, and more specifically of the complex
regime of matter pressure ionization and metallization.

\begin{acknowledgments}
We are very grateful to Alexander Potekhin and G\'erard Massacrier 
for very useful discussions and insightful remarks.
\end{acknowledgments}

\bibliographystyle{apsrev}
\bibliography{references.bib}

\end{document}